\documentclass[]{pasj02} 
\usepackage[switch,mathlines]{lineno} 
\usepackage{graphicx}

\jyear{2026}
\Received{}
\Accepted{}

\newcommand{\rev}[1]{\textcolor{black}{#1}}
\newcommand{\revsec}[1]{\textcolor{black}{#1}}
\newcommand{\revthird}[1]{\textcolor{black}{#1}}
\newcommand{\revfourth}[1]{\textcolor{black}{#1}}

\begin{document} 

\title{Formation of Micron-sized Presolar Graphite Grains in Low-energy Supernova Ejecta}

\author{
 Kengo \textsc{Shinoda},\altaffilmark{1,2}\altemailmark\orcid{0009-0006-0552-9925} \email{kengo0317@g.ecc.u-tokyo.ac.jp} 
 Ryo \textsc{Tazaki},\altaffilmark{1}\orcid{0000-0003-1451-6836}
 Takato \textsc{Tokuno},\altaffilmark{3, 1}\orcid{0000-0003-1579-5937}
 and 
 Yudai \textsc{Suwa}\altaffilmark{1,4}\orcid{0000-0002-7443-2215}
}
\altaffiltext{1}{Department of Earth Science and Astronomy, The University of Tokyo, Tokyo 153-8902, Japan}
\altaffiltext{2}{Department of Astronomy, The University of Tokyo, Tokyo 113-0033, Japan}
\altaffiltext{3}{Department of Earth and Space Science, The University of Osaka, Osaka 560-0043, Japan
}
\altaffiltext{4}{Center for Gravitational Physics and Quantum Information, Yukawa Institute for Theoretical Physics, Kyoto University, Kyoto 606-8502, Japan
}



\maketitle

\begin{abstract}
Presolar graphite grains with supernova isotopic signatures provide direct evidence that large \revsec{graphite} grains formed in supernova ejecta. However, theoretical calculations of dust formation in supernovae have typically predicted \revsec{graphite} grains smaller than $0.1\,{\rm \mu m}$, leaving unclear the physical conditions required to form \revsec{those} grains with radii of order $1\,{\rm \mu m}$. 
We systematically investigate graphite nucleation and growth in expanding supernova ejecta
\revsec{by varying the explosion energy, ejecta mass, initial ejecta radius, and ejected $^{56}$Ni mass. We use}
 a simplified one-zone model that follows the temperature and density evolution including adiabatic cooling, radiative cooling, and heating by radioactive nickel. We show that the final graphite grain size is mainly determined by
\revsec{the product of the mass density of atomic carbon available
for grain growth and the local cooling timescale, both evaluated
at a reference temperature of $2000\,{\rm K}$.}
\rev{Our models show that graphite growth is favored by lower explosion energy, larger ejecta mass, and smaller initial ejecta radius, which keep the gas denser near the graphite-condensation temperature. When radioactive heating and radiative cooling are included, explosions with $E_{\rm exp}\sim10^{50}\,{\rm erg}$ and $M_{\rm Ni}\lesssim10^{-3}\,M_\odot$ can form $1$--$10\,\mu{\rm m}$ graphite grains in the ejected helium layer or carbon--oxygen core, whereas $E_{\rm exp}\sim10^{51}\,{\rm erg}$ models do not. These results suggest that micron-sized presolar supernova graphite preferentially forms in faint-supernova-like explosions with low explosion energies, little ejected nickel, and helium layer or carbon--oxygen core ejecta.}
\end{abstract}


\section{Introduction}
\label{sec:Intro}
\rev{In supernovae,} newly formed or pre-existing dust can produce infrared excesses, optical attenuation and reddening, and asymmetric emission-line profiles 
\citep{Matsuura2011Sci, Gall2014Natur,Wesson2015MNRAS, Bevan2016MNRAS, Sarangi2018SSRv, Davis2026arXiv}. 
Such dust signatures play an important role in interpreting their light curves and spectral energy distributions.
The observational impact of dust depends not only on the total dust mass but also sensitively on the grain size distribution. 
\revsec{The grain size can affect the absorption and scattering properties, particularly when the grain size is comparable to or larger than the relevant wavelength.}

\revsec{Grain size also affects the survival of dust against
reverse shocks and subsequent shocks driven by other supernovae.}
Indeed, sputtering calculations for reverse shocks have shown that small grains with $a_{\rm ini}\lesssim 0.05$--$0.1\,{\rm\mu m}$ are efficiently destroyed, whereas grains with $a_{\rm ini}\gtrsim 0.1$--$0.2\,{\rm \mu m}$ are more likely to survive 
\citep{Nozawa2007ApJ, Silvia2010ApJ}.
In addition to this size dependence, the survival fraction in actual supernova remnants also depends on the initial grain-size distribution, dust composition, ejecta morphology, and ambient environment 
\citep{Micelotta2016A&A}. 

\revsec{Constraining grain-size distributions from astronomical
observations is challenging because the inferred distributions
are generally model dependent and can be degenerate with other
dust and radiative-transfer parameters
\citep{Zubko2004ApJS,Wesson2015MNRAS}. In contrast, laboratory measurements of individual presolar
grains preserved in meteorites directly determine their sizes,
compositions, and microstructures, while their isotopic ratios
provide clues to their astrophysical origins}
\citep{Zinner2014mcp, Nittler2016ARA&A}.
Among these diagnostics, excesses of $^{44}{\rm Ca}$ produced by the decay of $^{44}{\rm Ti}$ provide a representative signature of a supernova origin and have been reported for graphite grains and SiC grains from the Murchison meteorite 
\citep{Nittler1996ApJ, Nittler2016ARA&A}. 
\revsec{Presolar grains have been identified in a variety of mineral phases. Examples include silicates such as $\mathrm{Mg_2SiO_4}$ and $\mathrm{MgSiO_3}$ (e.g., \cite{Nguyen2010ApJ}), as well as alumina ($\mathrm{Al_2O_3}$) (e.g., \cite{Stroud2004Sci}). Here, we focus on graphite ($\mathrm{C}$), whose single-element composition provides a simple case for modeling nucleation and grain growth.} 
Presolar graphite grains identified as supernova condensates can reach sizes of $\sim1\,{\rm \mu m}$ or larger 
\citep{Amari1994GeCoA, Amari2014GeCoA, Nittler2016ARA&A}, providing an important benchmark for graphite formation in supernova ejecta.

However, existing theoretical calculations of dust formation in supernova ejecta typically predict much smaller \revsec{graphite} grains. The characteristic sizes of \revsec{graphite} grains are of order several tens of nanometers, 
\revsec{with}
the largest grains generally \revsec{being smaller than} $\sim0.1\,{\rm \mu m}$ 
\citep{Todini2001MNRAS, Nozawa2003ApJ, Sarangi2015A&A}. 
These \revsec{grain sizes} are substantially smaller than \revsec{those of} presolar graphite \revsec{grains}. 

\revsec{This discrepancy may arise from the limited range of supernova properties in previous calculations.}
Many previous calculations of dust formation in core-collapse supernovae have adopted ejecta models representative of ordinary type II supernovae, 
\revsec{characterized by explosion energies around $10^{51}\,{\rm erg}$, ejecta masses around $10\,M_\odot$, and synthesized $^{56}{\rm Ni}$ masses of $0.01$--$0.1\,M_\odot$} 
\citep{Todini2001MNRAS, Nozawa2003ApJ, Sarangi2015A&A}.
\revsec{However,} supernovae span a broad range of explosion energies, ejecta masses, and synthesized $^{56}{\rm Ni}$ masses.
Even among ordinary core-collapse supernovae, typical inferred ranges are \revsec{explosion energies of} $\sim10^{50}$--$10^{51}\,{\rm erg}$, \revsec{ejecta masses of} $\sim$ a few--$10M_\odot$, and \revsec{synthesized $^{56}{\rm Ni}$ masses of} $\sim10^{-3}$--$10^{-1}M_\odot$ 
\citep{Muller2017ApJ, Martinez2022A&A}.
At the lower-energy end, faint supernovae and failed-supernova candidates extend this diversity toward explosions with very small $^{56}{\rm Ni}$ production and weak mass ejection 
\citep{Hamuy2003ApJ, Pejcha2015ApJ, Adams2017MNRAS, Taddia2018A&A, Martinez2022A&A, De2026ApJ, De2026Sci}.
These macroscopic parameters affect the temperature and density evolution of the ejecta through expansion, density dilution, radioactive heating, and radiative cooling, and therefore directly influence the conditions under which dust forms and grows.
The dependence of dust formation on ejecta mass and expansion velocity has also been investigated for kilonova ejecta \citep{Domoto2026arXiv}.
\revsec{Despite this diversity,} how the final grain size changes over a broader parameter space, particularly toward lower explosion energies and smaller ejected $^{56}{\rm Ni}$ masses, has not been systematically explored.

In this paper, 
\rev{we aim to study graphite dust formation under various conditions for supernova explosion. To this end, we develop a simple one-zone model to treat dust formation in supernova ejecta to facilitate an efficient parameter study.}
Our main goal is to clarify how the explosion energy, ejecta mass, ejected $^{56}{\rm Ni}$ mass, and ejecta layer \revsec{affect} the final graphite grain size through the thermal history of the ejecta. 
\rev{We find that micron-sized graphite preferentially forms in low-energy, nickel-poor explosions with relatively large ejecta masses and ejected helium layer or carbon--oxygen core material, where the gas remains sufficiently dense near the graphite-condensation temperature.}

\rev{This paper is organized as follows. In Section~\ref{sec:onezone_model}, }
we describe the one-zone ejecta model and the treatment of its thermal evolution. In Section~\ref{sec:dust_formation}, we present the calculation of graphite nucleation and grain growth based on classical nucleation theory. In Section~\ref{sec:results}, we examine the conditions under which micron-sized graphite can form, using both Ni-free benchmark models and a parameter survey including $^{56}{\rm Ni}$ heating. In Section~\ref{sec:discussion}, we discuss the implications of our results for low-energy supernovae and the origin of presolar supernova graphite. Finally, we summarize our conclusions in Section~\ref{sec:conclusion}.

\section{One-zone model of supernova ejecta}
\label{sec:onezone_model}

To calculate graphite nucleation and grain growth in supernova ejecta, we first consider the time evolution of the gas density and temperature.
\rev{The resulting density--temperature trajectories determine the supersaturation ratio, the nucleation rate, and the subsequent growth rate of graphite grains, and are used as input for the dust-formation calculation described in Section~\ref{sec:dust_formation}.}

In this study, we use a simplified one-zone ejecta model.
We assume that the ejecta is spherical and spatially homogeneous, with volume $V=4\pi R^3/3$ and density $\rho=3M_{\rm ej}/(4\pi R^3)$, where $R$ is the characteristic ejecta radius.
The model is specified by \revsec{the initial explosion energy $E_{\rm exp}$, ejecta mass $M_{\rm ej}$, initial $^{56}\mathrm{Ni}$ mass $M_{\rm Ni,0}$, initial radius $R_0$, and initial internal-energy fraction $f_{\rm int}$. We define $t=0$ as the initial epoch of the one-zone evolution and $E_{\rm exp}\equiv E_{\rm int}(0)+E_{\rm kin}(0)$, where $E_{\rm int}$ and $E_{\rm kin}$ are the internal and kinetic energies of the ejecta, respectively. The initial conditions are}
    $R(0)=R_0,\  E_{\rm int}(0)=f_{\rm int}E_{\rm exp},\ E_{\rm kin}(0)=(1-f_{\rm int})E_{\rm exp}.$

\revsec{At a given radius $R$, the internal energy determines the temperature and pressure through an equation of state that includes radiation and ideal-gas contributions, 
\begin{align}
    \varepsilon = a T^4 + \frac{3}{2}\frac{\rho}{\revthird{\bar{\mu}}m_p}k_BT,
    \label{eq:eos_idrad_intene}
\end{align}
where $\varepsilon \equiv E_{\rm int}/V,\,a,\, m_p,\, k_B,\, \revthird{\bar{\mu}}$ are the internal energy density, radiation constant, proton mass, Boltzmann constant, and \revthird{mean molecular weight defined by the total number density of ions and electrons,} respectively.
The pressure is then calculated as
\begin{align}
    p = \frac{aT^4}{3} + \frac{\rho}{\revthird{\bar{\mu}}m_p}k_B T.
    \label{eq:eos_idrad_pres}
\end{align}
}
\revthird{We adopt a fixed mean molecular weight of
$\bar{\mu}=1$ in all calculations.}

The thermal evolution of the ejecta is calculated by solving the energy equation including radioactive heating, radiative cooling, and adiabatic expansion
\revsec{\citep{Arnett1980ApJ,Arnett1982ApJ}}. 
The internal energy evolves as
\begin{align}
    \frac{dE_{\rm int}}{dt} &= - L + \dot{Q} - p\frac{dV}{dt},
    \label{eq:derivation_of_Eint} 
\end{align}
where $L$ and $\dot{Q}$ are the radiative luminosity and radioactive heating rate.

\rev{We first describe the radiative luminosity $L$ in Eq.~(\ref{eq:derivation_of_Eint}). In the optically thick limit, the corresponding diffusion luminosity is written as}
\begin{align}
    \rev{
    L_{\rm diff}
    = \frac{E_{\rm rad}}{t_{\rm diff}},
    }
    \label{eq:luminosity_diffusion_approx}
\end{align}
\rev{where $E_{\rm rad}=aT^4V$ is the radiation energy and $t_{\rm diff}$ is the diffusion timescale.}
\rev{For the one-zone ejecta, the diffusion timescale is}
\begin{align}
    \rev{
    t_{\rm diff}
    = \frac{\tau R}{c},
    }
    \label{eq:diffusion_timescale}
\end{align}
\revsec{where $c$ is the speed of light and $\tau$ is the optical depth, }
\begin{align}
    \tau = \kappa \rho R
    = \frac{3\kappa M_{\rm ej}}{4\pi R^2}.
    \label{eq:tau_opt}
\end{align}
\revsec{Here, $\kappa$ denotes the Rosseland-mean opacity.}
\rev{In the optically thin limit, photons escape on the timescale}
\begin{align}
    \rev{
    t_{\rm lc} = \frac{R}{c},
    }
    \label{eq:light_crossing_timescale}
\end{align}
\rev{where $t_{\rm lc}$ is the light-crossing timescale.}
\rev{We interpolate between the optically thick and thin limits by taking the effective photon escape timescale to be $t_{\rm diff}+t_{\rm lc}$. The radiative luminosity is then}
\begin{align}
    \rev{
    L
    = \frac{E_{\rm rad}}{t_{\rm diff}+t_{\rm lc}}
    = \frac{\tau}{1+\tau}L_{\rm diff},
    }
    \label{eq:L_tau}
\end{align}
\rev{which approaches $L = L_{\rm diff}$ in the optically thick limit and $L = E_{\rm rad}/t_{\rm lc}$ in the optically thin limit 
\citep{Metzger2014MNRAS, Metzger2017MNRAS,Chen2024MNRAS}.}
In this work, we adopt \revsec{a constant electron-scattering opacity}, $\kappa = 0.2\,{\rm cm^2\,g^{-1}}$
for all models with radiative cooling\revsec{, as commonly adopted in simplified supernova diffusion models \citep{Kasen2010ApJ, Metzger2014MNRAS,Metzger2015MNRAS}}. 
For comparison, we also consider a purely adiabatic model by setting
$L=0$, 
which isolates the effect of $pdV$ work and radioactive heating on the thermal evolution of the ejecta.

\revsec{We next describe the radioactive heating term
$\dot{Q}$ in Eq.~(\ref{eq:derivation_of_Eint}).
The $^{56}{\rm Ni}\rightarrow{}^{56}{\rm Co}
\rightarrow{}^{56}{\rm Fe}$ decay chain releases energy
in $\gamma$-rays, positrons, and neutrinos
\citep{Nadyozhin1994ApJS}.
We neglect neutrino energy deposition and distinguish
between the $\gamma$-ray and positron kinetic-energy
contributions to the deposited heating rate:}
\begin{align}
    \dot{Q}
    = f_{\rm dep}(\tau_\gamma)\,\dot{Q}_{\gamma}
    + \dot{Q}_{\rm pos}.
    \label{eq:heating_deposition_split}
\end{align}
\revsec{We adopt the radioactive energy-release rates
for the
$^{56}{\rm Ni}\rightarrow{}^{56}{\rm Co}\rightarrow{}^{56}{\rm Fe}$
decay chain given by \citet{Sharon2020MNRAS}:}
\begin{equation}
    \revsec{
    \begin{aligned}
        \dot{Q}_{\gamma}(t)
        &=
        \frac{M_{\rm Ni,0}}{M_\odot}
        \left[
            6.54\,e^{-t/\tau_{\rm Ni}}
            + 1.38\,e^{-t/\tau_{\rm Co}}
        \right]
        \times 10^{43}\,{\rm erg\,s^{-1}},
        \\
        \dot{Q}_{\rm pos}(t)
        &=
        4.64\,\frac{M_{\rm Ni,0}}{M_\odot}
        \left[
            e^{-t/\tau_{\rm Co}}
            - e^{-t/\tau_{\rm Ni}}
        \right]
        \times 10^{41}\,{\rm erg\,s^{-1}},
    \end{aligned}
    }
    \label{eq:Qgamma_Qpos}
\end{equation}
\revsec{where $M_{\rm Ni,0}$ is the initial $^{56}{\rm Ni}$
mass, and $\tau_{\rm Ni}=8.76\,{\rm d}$ and
$\tau_{\rm Co}=111.4\,{\rm d}$ are the mean lifetimes.
The $\gamma$-ray component includes photons produced by
positron annihilation, whereas $\dot{Q}_{\rm pos}$
contains only the positron kinetic energy.
We assume that this kinetic energy is deposited locally
and instantaneously.}
\revsec{To describe incomplete $\gamma$-ray trapping
in our one-zone model, we approximate the deposited
fraction as}
\begin{align}
    f_{\rm dep}(\tau_\gamma)=1-e^{-\tau_\gamma}.
    \label{eq:fdep}
\end{align}
We parameterize the $\gamma$-ray optical depth as
\begin{align}
    \tau_\gamma
    = \kappa_\gamma \rho R
    = \frac{3\kappa_\gamma M_{\rm ej}}{4\pi R^2},
    \label{eq:tau_gamma}
\end{align}
where $\kappa_\gamma$ is the effective (gray)
$\gamma$-ray opacity.
\revsec{We adopt a constant value of
$\kappa_\gamma=0.027\,{\rm cm^2\,g^{-1}}$,
which is reasonable in the $\gamma$-ray transport in supernova ejecta \citep{Sutherland1984ApJ, Maeda2003ApJ}.}

\rev{Finally, we describe the dynamical evolution associated with the $p\,dV$ work in Eq.~(\ref{eq:derivation_of_Eint}). The kinetic energy evolves as}
\begin{align}
    \frac{dE_{\rm kin}}{dt} &= p \frac{dV}{dt},
    \label{eq:derivation_of_Ekin}
\end{align}
\rev{where $E_{\rm kin}$ is the kinetic energy of the ejecta. The ejecta radius then evolves as}
\begin{align}
    \rev{
    \frac{dR}{dt}
    = \left(\frac{2E_{\rm kin}}{M_{\rm ej}}\right)^{\frac{1}{2}}.
    }
    \label{eq:radius_evolution}
\end{align}
\revsec{We simultaneously solve Eq.~(\ref{eq:derivation_of_Eint}), Eq.~(\ref{eq:derivation_of_Ekin}), and Eq.~(\ref{eq:radius_evolution}) as three coupled ordinary differential equations for $E_{\rm int}$, $E_{\rm kin}$, and $R$.}

\begin{figure}
    \includegraphics[width=1.\linewidth]{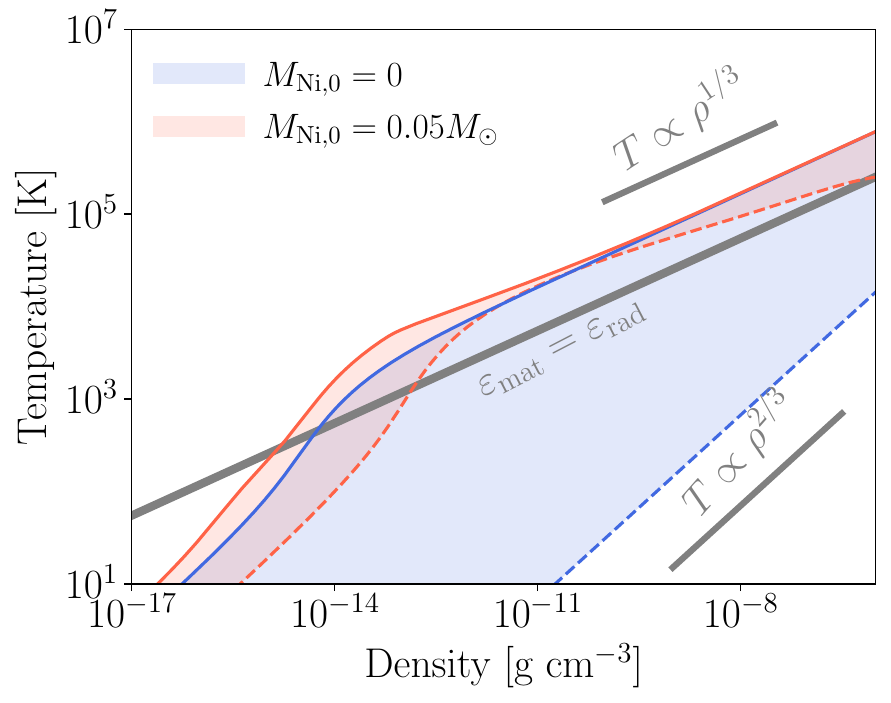}
    \caption{\rev{The density--temperature trajectories for representative
combinations of explosion energy and initial $^{56}$Ni mass.
Colors indicate $M_{\rm Ni,0}$, while line styles indicate $E_{\rm exp}$ (solid lines: $10^{51}\, {\rm erg}$, \revsec{dashed} lines: $10^{48}\, {\rm erg}$).
The gray \revsec{thick} line indicates $\varepsilon_{\rm mat}=\varepsilon_{\rm rad}$,
where $\varepsilon_{\rm rad}=aT^4$ and
$\varepsilon_{\rm mat}=3\rho k_B T / 2 m_p$ are the radiation
and ideal-gas contributions to the internal energy density, respectively.
\revsec{
The short gray line segments indicate the adiabatic scalings $T\propto\rho^{1/3}$ and $T\propto\rho^{2/3}$ in the radiation- and matter-dominated limits, respectively.
}
We use $f_{\rm int}=0.1$, $M_{\rm ej}=10\,M_\odot$, and
$R_0=10^{13}\,{\rm cm}$ for all models.}}
    \label{fig:rho-T_plane}
\end{figure}

Fig.~\ref {fig:rho-T_plane} shows the resulting density--temperature
trajectories for representative combinations of explosion parameters. 
\revsec{At fixed $M_{\rm ej}$, $R_0$, and $f_{\rm int}$, increasing $E_{\rm exp}$ from $10^{48}$ to $10^{51}\,{\rm erg}$ raises the initial temperature and shifts the trajectory toward higher temperatures at a given density. Adding $M_{\rm Ni,0}=0.05\,M_\odot$ keeps the $10^{48}\,{\rm erg}$ ejecta hotter during expansion, so they reach temperatures near condensation ($\sim 1000 \,{\rm K}$) at lower densities. The effect of radioactive heating is much smaller for $E_{\rm exp}=10^{51}\,{\rm erg}$.}

\rev{The gray line indicates where the radiation and ideal-gas
contributions to the internal energy in Eq.~(\ref{eq:eos_idrad_intene})
are equal; the radiation contribution dominates above this line,
whereas the ideal-gas contribution dominates below it.}
\rev{For reference, adiabatic expansion gives
$T\propto\rho^{1/3}$ in the radiation-dominated limit and
$T\propto\rho^{2/3}$ in the matter-dominated limit, as indicated
by the \revsec{short gray line} segments in Fig.~\ref {fig:rho-T_plane}.}
\rev{Some trajectories evolve from the radiation-dominated to the
matter-dominated regime during expansion, whereas others are already
matter dominated from the beginning of the evolution shown here.}

\section{Dust Nucleation and Grain Growth}
\label{sec:dust_formation}

In this study, dust formation in supernova ejecta is treated as a two-step process consisting of nucleation and subsequent grain growth. 
\rev{We calculate these processes along the density and temperature
evolution obtained in Section~\ref{sec:onezone_model}, so that the
thermal history of the ejecta directly determines the conditions for
dust nucleation and grain growth.}
The basic framework of the dust formation calculation follows \citet{Nozawa2003ApJ}, and is formulated using steady-state classical nucleation theory under the key species approximation.

\subsection{Overview of nucleation theory}

In cooling and expanding ejecta, atoms and molecules in the gas phase form condensation nuclei, which subsequently grow into dust grains.
For each dust species $j$, the nucleation rate is described by the steady-state nucleation rate $J_j$.
\rev{Following \citet{Nozawa2003ApJ}, we formulate nucleation and grain growth
using the key species approximation introduced by \citet{Kozasa1987PThPh}.}
\rev{Under this approximation, both the nucleation and grain-growth rates are assumed to be controlled by the least abundant colliding gas species in the reaction network.}
\rev{Although the formalism is written for a general dust species $j$, we consider
only graphite in this study, for which atomic carbon is the key species.}

Under this approximation, the steady-state nucleation rate for dust species $j$ is given by
\begin{align}
J_j =
\alpha_{j} \Omega_{1j} c_{1j}^2
\sqrt{\frac{2\zeta_j}{\pi m_{1j}}}
\exp\left[
-\frac{4}{27}\frac{\mu_j^3}{(\ln S_j)^2}
\right],
\label{eq:nucleation_rate}
\end{align}
where $\alpha_{j}$, $\zeta_j$, and $S_j$ denote the sticking probability, surface free energy, and supersaturation ratio of the key species for dust species $j$, respectively. \rev{We assume $\alpha_j=1$ throughout this study.}
Here, $\Omega_{1j} \equiv 4\pi a_{1j}^3/3$ is the volume of a monomer, while $c_{1j}$, $m_{1j}$, and $a_{1j}$ represent the number density \rev{in the gas phase}, mass, and atomic radius of the key species, respectively.
The dimensionless surface tension is defined as
\begin{align}
\mu_j \equiv \frac{4\pi a_{1j}^2 \zeta_j}{k_B \rev{T}},
\end{align}
\rev{where we have assumed that the temperature of the condensed phase is
equal to the gas temperature.}

The supersaturation ratio $S_j$ is defined by taking into account the chemical reactions relevant to dust formation as
\begin{align}
\ln S_j
=
-\frac{\Delta G_j^0}{k_BT}
+
\sum_i \nu_{ij} \ln P_{ij},
\end{align}
where $\Delta G_j^0$ is the standard Gibbs free energy of formation per key species molecule, $P_{ij}$ denotes the partial pressure of reactants and products, and $\nu_{ij}$ is the stoichiometric coefficient.
As the ejecta cools and expands, nucleation begins once $S_j$ exceeds unity.
\revsec{We set $J_j=0$ for $S_j\leq1$ and evaluate Eq.~(\ref{eq:nucleation_rate}) only for $S_j>1$.}

\subsection{Grain growth and depletion of key species}

After nucleation, dust grains are assumed to grow by surface accretion through collisions with key species.
The growth rate of the grain size $r_j$ is proportional to the number density of the key species and is given by
\begin{align}
\label{eq:grain_growth_rate}
\frac{dr_j}{dt}
=
\alpha_{j}\,\Omega_{1j}\,c_{1j}\,v_{\rm th, {\it j}},
\end{align}
where
\begin{align}
v_{\rm th, {\it j}} = \sqrt{\frac{k_BT}{2\pi m_{1j}}}
\end{align}
is the thermal velocity of the key species.

\rev{To account for the depletion of the key species, we introduce
the normalized monomer abundance, }
\begin{align}
    \rev{
    Y_j \equiv \frac{c_{1j}}{\tilde{c}_{1j}},
    }
    \label{eq:Yj}
\end{align}
\rev{where $\tilde{c}_{1j}$ is the number density of the key species
in the absence of depletion by dust formation.}
The nucleation rate and grain growth are coupled self-consistently by solving the following integral equation:
\begin{align}
1 - 
\rev{Y_j}
=
\int_{t_S}^{t}
\frac{1}{\tilde{c}_{1j}}
J_j(t')
\frac{1}{\Omega_{1j}}
\frac{4\pi}{3}
r_j^3(t,t')\,dt'.
\label{eq:monomer_depletion}
\end{align}
Here, $r_j(t,t')$ represents the radius at time $t$ of a grain nucleated at time $t'$, and $t_S$ is the time at which $S_j$ first exceeds unity.

To describe the evolution of the grain population, we introduce the $\ell$-th moment as
\begin{align}
K_j^{(\ell)}
=
\int_{t_S}^{t}
\frac{J_j(t')}{\tilde{c}_{1j}(t')\Omega_{1j}}
\frac{4\pi}{3}
r_j^{\ell}(t,t')\,dt'.
\label{eq:moment_definition}
\end{align}
\rev{Differentiating Eq.~(\ref{eq:moment_definition}) gives}
\begin{align}
\frac{dK_j^{(\ell)}}{dt}
=
\frac{J_j}{\tilde{c}_{1j}}
\frac{4\pi}{3\Omega_{1j}}
r_{*j}^{\ell}
+
\ell K_j^{(\ell-1)}\frac{dr_j}{dt}
\qquad (\ell = 1,2,3),
\label{eq:moment_evolution}
\end{align}
where
\begin{align}
r_{*j} \equiv \frac{2a_{1j}\mu_j}{3\ln S_j}
\end{align}
is the critical radius required for a stable nucleus.
\rev{During graphite nucleation in our models,
$r_{*j}$ is typically of order $10^{-8}$--$10^{-7}\,{\rm cm}$,
corresponding to a few monomer radii.}
\rev{For the zeroth moment, Eq.~(\ref{eq:moment_definition}) gives}
\begin{align}
    \rev{
    \frac{dK_j^{(0)}}{dt}
    =
    \frac{J_j}{\tilde{c}_{1j}}
    \frac{4\pi}{3\Omega_{1j}}.
    }
    \label{eq:moment_zeroth}
\end{align}

\rev{For graphite, Eqs. (\ref{eq:monomer_depletion}) and
(\ref{eq:moment_definition}) give}
\begin{align}
    \rev{
    \frac{dY_j}{dt}
    = -\frac{dK_j^{(3)}}{dt},
    }
    \label{eq:Yj_evolution}
\end{align}
\rev{with $Y_j(0)=1$.}

\rev{We simultaneously integrate
Eqs. (\ref{eq:grain_growth_rate}),
 (\ref{eq:moment_evolution}),
 (\ref{eq:moment_zeroth}),
 and (\ref{eq:Yj_evolution})
for $r_j$, $K_j^{(0)}$--$K_j^{(3)}$, and $Y_j$,
using $c_{1j}=Y_j\tilde{c}_{1j}$ at each time step.
The mass-averaged grain size $r_{{\rm gr},j}$ and the grain number density
$n_{{\rm gr},j}$ are then obtained from the moments as}
\begin{align}
r_{{\rm gr},j}
=
\left(
\frac{K_j^{(3)}}{K_j^{(0)}}
\right)^{1/3},
\end{align}
and
\begin{align}
n_{{\rm gr},j}
=
a_{1j}^3 K_j^{(0)} \tilde{c}_{1j}.
\end{align}

\subsection{Computational setup and simplifications}

In this study, we focus exclusively on the formation of graphite and do not consider the formation of silicate or other dust species.
In chemically mixed ejecta with solar abundance ratios ($\mathrm{C/O} < 1$), carbon is expected to be predominantly locked into CO molecules, leaving little free carbon available for graphite formation.

The number density of carbon key species contributing to graphite formation is parameterized as
\begin{align}
\tilde{c}_{1\mathrm{C}}
=
(1-f_{\mathrm{CO}})
\frac{X_{\mathrm{C}}\rho}{m_{1{\rm C}}},
\end{align}
where \revsec{$X_{\rm C}$ and} $f_{\mathrm{CO}}$ \revsec{are the carbon mass fraction in the ejecta} and the fraction of carbon locked into CO molecules, respectively, \rev{and $m_{1 {\rm C}} \equiv 12 m_p$ is the carbon atom mass}.

Rather than assuming complete CO formation ($f_{\mathrm{CO}} = 1$) for $\mathrm{C/O} < 1$, we treat the effective carbon abundance
\begin{align}
X'_{\mathrm{C}} \equiv (1-f_{\mathrm{CO}}) X_{\mathrm{C}}
\end{align}
as a free parameter.
\rev{This parametrization allows us to systematically explore the impact of
incomplete CO formation on graphite nucleation without explicitly modeling
the non-equilibrium CO chemistry, including CO destruction by energetic
particles produced by radioactive decay
\citep{Clayton1999Sci,Cherchneff2009ApJ,Sarangi2013ApJ}.}

\rev{For graphite, we adopt the monomer radius of
$a_{1 {\rm C}}=1.281\times10^{-8}\,{\rm cm}$, a surface free energy of
$\zeta_{\rm C}=1400\,{\rm erg\,cm^{-2}}$, and approximate the Gibbs free energy
for carbon condensation as $\Delta G^0_{\rm C}/(k_BT)=-A/T+B$, with
$A=8.64726\times10^4\,{\rm K}$ and $B=19.0422$, following
\citet{Nozawa2003ApJ}.}

\revsec{We use the density and temperature histories obtained
in Section~\ref{sec:onezone_model} as input to the
dust-formation calculation.
The density $\rho(t)$ sets the undepleted carbon number
density, $\tilde{c}_{1{\rm C}}=X'_{\rm C}\rho/m_{1{\rm C}}$,
while monomer depletion gives
$c_{1{\rm C}}=Y_{\rm C}\tilde{c}_{1{\rm C}}$.
The temperature $T(t)$ determines $\mu_{\rm C}$ and
$v_{{\rm th},{\rm C}}$ and, together with the carbon
partial pressure, determines $\ln S_{\rm C}$.
These quantities enter the nucleation and grain-growth
rates through Eqs.~(\ref{eq:nucleation_rate})
and (\ref{eq:grain_growth_rate}), respectively.}

\section{Results}
\label{sec:results}

\subsection{\revsec{Time evolution and final sizes of graphite grains}}
\begin{figure*}
    \includegraphics[width=1.0\linewidth]{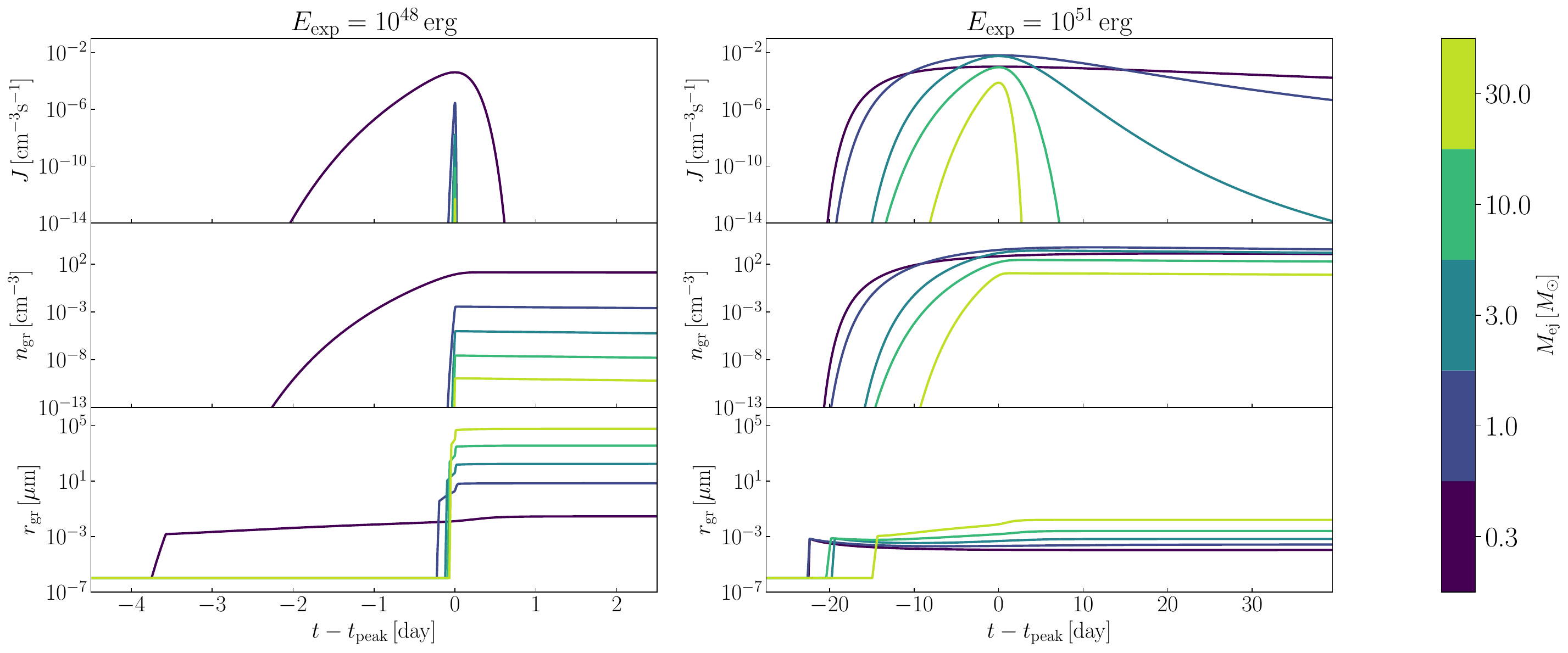}
    \caption{
    Time evolution of the steady-state nucleation rate $J$, \rev{grain number density $n_{\rm gr}$, and mass-averaged graphite grain size $r_{\rm gr}$} for different ejecta masses $M_{\rm ej}$. 
    The left and right columns show models with $E_{\rm exp}=10^{48}\,{\rm erg}$ and \revsec{$10^{51}\,{\rm erg}$}, respectively. In each model, the time is measured relative to the peak of the nucleation rate, $t_{\rm peak}$. 
    The top, middle, and bottom panels show $J$, \rev{$n_{\rm gr}$, and $r_{\rm gr}$,} respectively. The color bar indicates $M_{\rm ej}$. 
    \revsec{All models use $f_{\rm int}=0.9$, $R_0=10^{12}\,{\rm cm}$, and $X'_{\rm C}=3\times10^{-3}$.}
    }
    \label{fig:time-tcond_dustval}
\end{figure*}

\begin{table*}[t]
\caption{Model summary for $R_0=10^{12}\,{\rm cm}$, $f_{\rm int}=0.9$, and $X'_{\rm C}=3\times10^{-3}$.}
\label{tab:Jspeak_summary}

{\centering
\begin{tabular}{ccccccc}
\hline
$E_{\rm exp}$ [erg]
& $M_{\rm ej}$ [$M_\odot$]
& $t_{\rm peak}$ [day]
& $r_{\rm gr,final}$ [${\rm \mu m}$]
& $\rho_{\rm peak}$ [${\rm g\,cm^{-3}}$]
& $T_{\rm peak}$ [K]
& $\tau_{\rm cool,peak}$ [day]
\\[2pt]
\hline

$10^{48}$ & $0.3$ & $110$ & $2.8 \times 10^{-2}$ & $9.4 \times 10^{-13}$ & $1900$ & $180$ \\
 & $1$   & $19$  & $6.8$                  & $3.4 \times 10^{-9}$  & $2400$ & $24$ \\
 & $3$   & $13$  & $170$                  & $1.4 \times 10^{-7}$  & $2700$ & $13$ \\
 & $10$  & $12$  & $3.5 \times 10^{3}$    & $3.6 \times 10^{-6}$  & $3100$ & $16$ \\
 & $30$  & $11$  & $5.8 \times 10^{4}$    & $6.8 \times 10^{-5}$  & $3500$ & $14$ \\

\hline

$10^{51}$ & $0.3$ & $79$  & $1.1 \times 10^{-4}$ & $7.4 \times 10^{-17}$ & $1200$ & $180$ \\
 & $1$   & $120$ & $2.6 \times 10^{-4}$ & $4.6 \times 10^{-16}$ & $1400$ & $260$ \\
 & $3$   & $170$ & $6.6 \times 10^{-4}$ & $2.5 \times 10^{-15}$ & $1500$ & $410$ \\
 & $10$  & $260$ & $2.5 \times 10^{-3}$ & $1.3 \times 10^{-14}$ & $1600$ & $640$ \\
 & $30$  & $300$ & $1.6 \times 10^{-2}$ & $1.3 \times 10^{-13}$ & $1800$ & $630$ \\

\hline
\end{tabular}
\par
}

\end{table*}

\begin{figure}[!t]
  {
  \centering

  \includegraphics[width=0.99\linewidth]{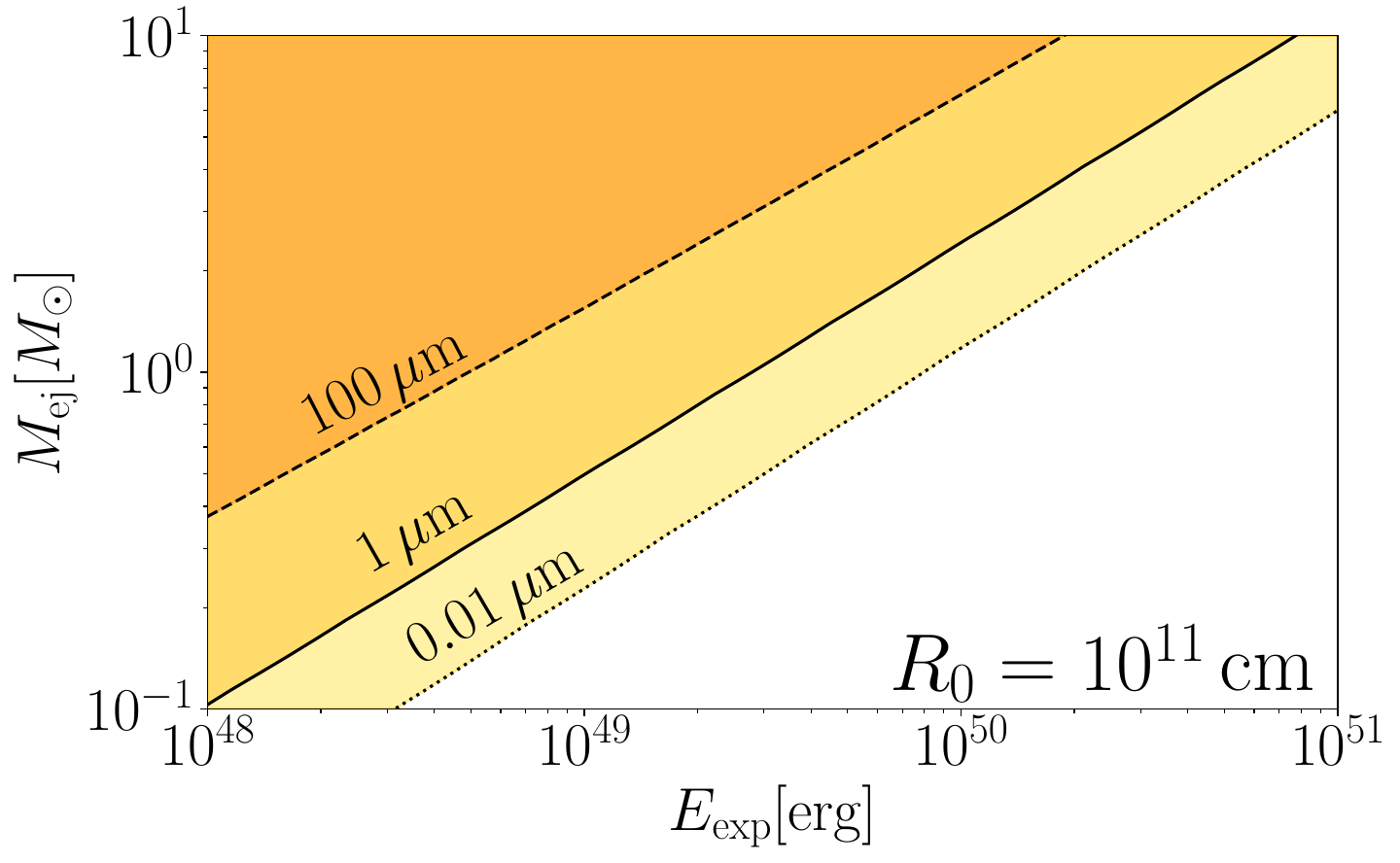}
  \vspace{-0.5cm}

  \includegraphics[width=0.99\linewidth]{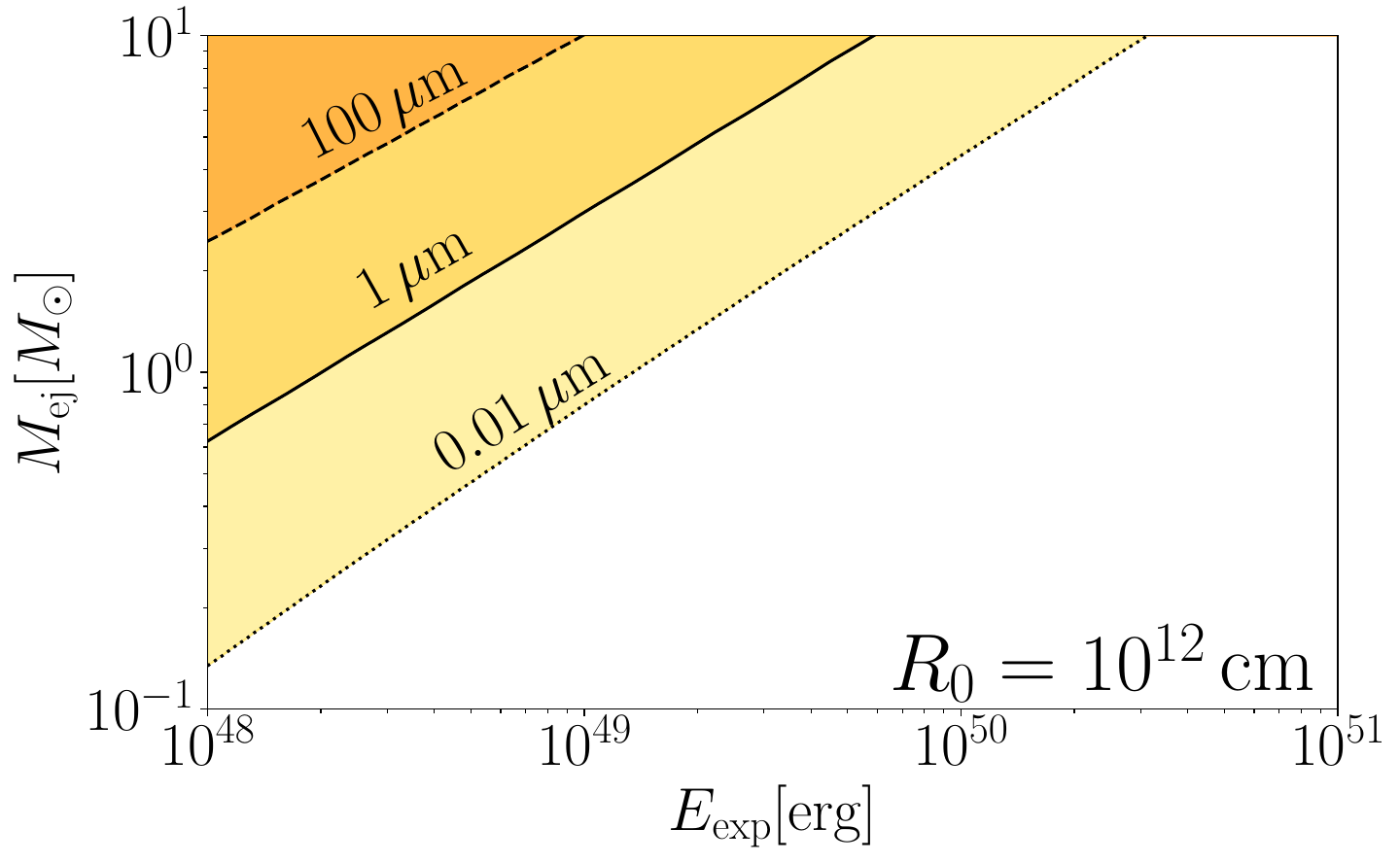}
  \vspace{-0.5cm}

  \includegraphics[width=0.99\linewidth]{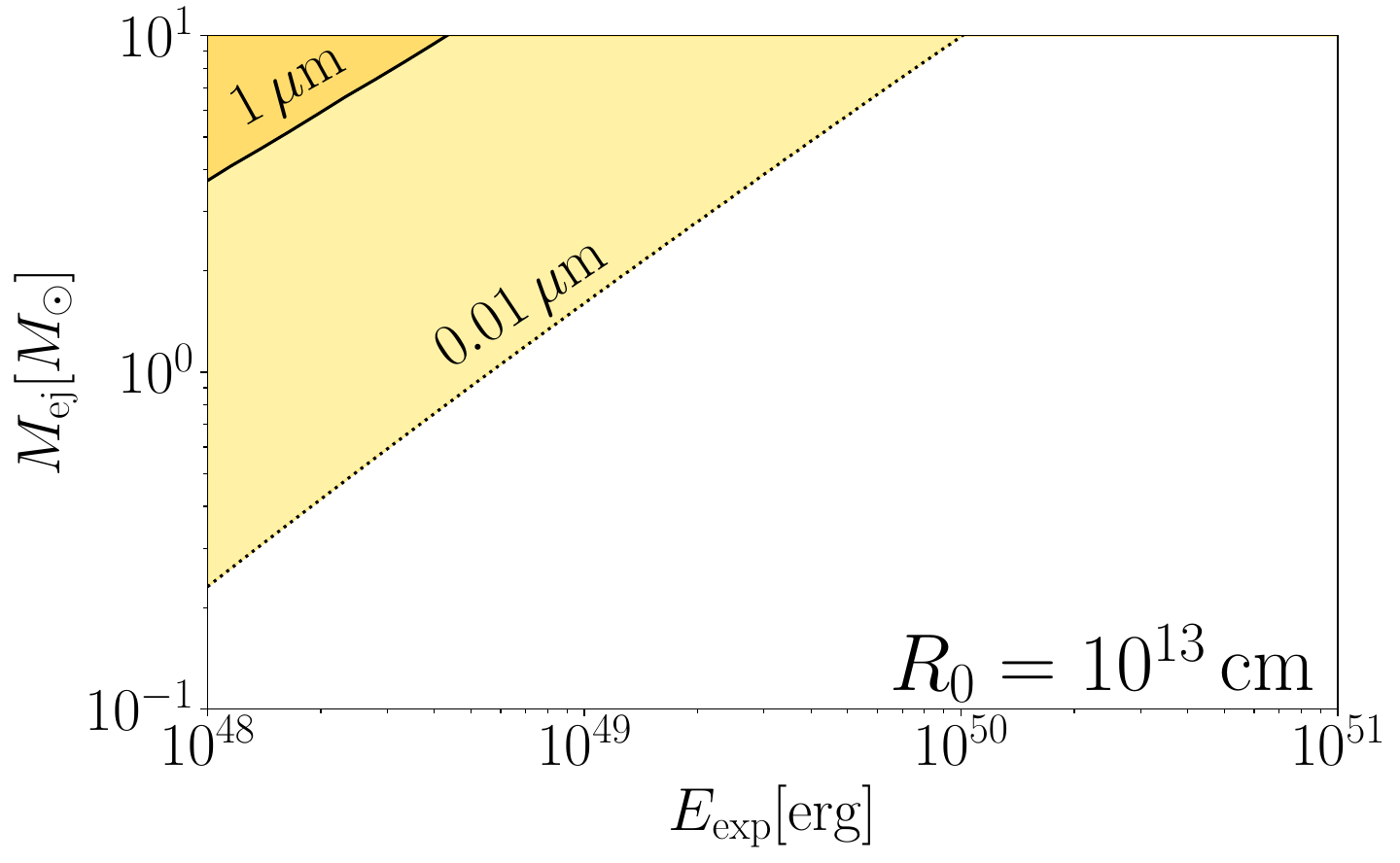}
  \vspace{-0.5cm}

  \includegraphics[width=0.99\linewidth]{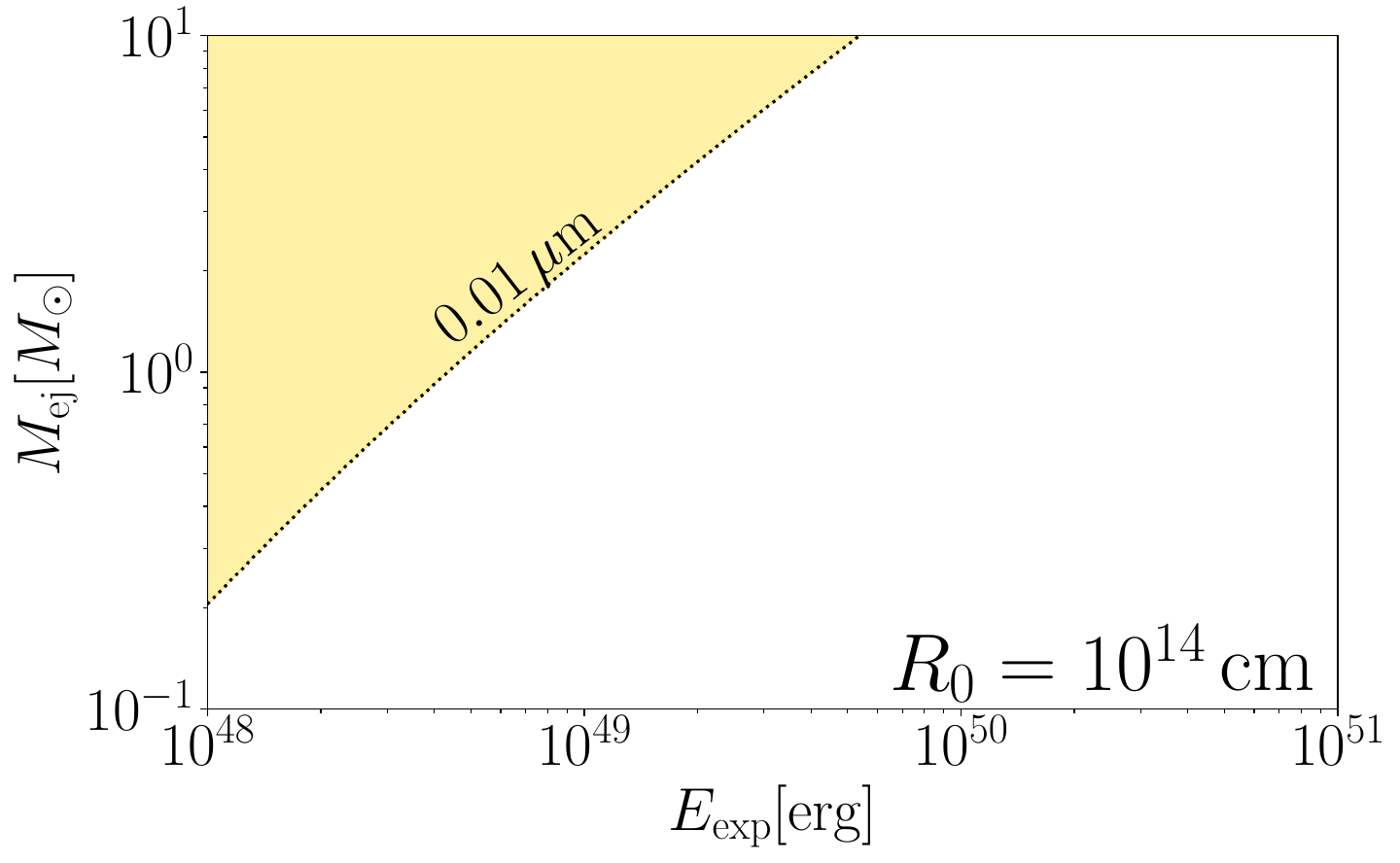}
  \par
  }

  \caption{
    Final mass-averaged dust radius on the $E_{\rm exp}$-$M_{\rm ej}$ plane
    \rev{for Ni-free ejecta with} four values of the initial radius $R_0$:
    $R_0 = 10^{11}$, $10^{12}$, $10^{13}$, and $10^{14}\,{\rm cm}$,
    shown from top to bottom, respectively.
    The horizontal axis is the explosion energy $E_{\rm exp}$, and the
    vertical axis is the ejecta mass $M_{\rm ej}$.
    The dotted, solid, and dashed lines indicate
    $r_{\rm gr}=0.01\,{\rm \mu m}$, $1\,{\rm \mu m}$, and
    $100\,{\rm \mu m}$, respectively.
    The shaded regions show the corresponding ranges separated by these contours.
  }
  \label{fig:Eex_Mej_R0_minipage}
\end{figure}

\revsec{To illustrate how the explosion energy and ejecta mass
affect graphite nucleation and growth, we first consider
Ni-free adiabatic models with
$E_{\rm exp}=10^{48}\,{\rm erg}$ and $10^{51}\,{\rm erg}$.
For simplicity, we omit radioactive heating and radiative
cooling here, although those effects will be considered in Section~\ref{subsec:Ni_heating}.}
\revsec{The colors indicate the ejecta mass,
which ranges from $0.3$ to $30\,M_\odot$
at each explosion energy.}

\revsec{We characterize graphite formation by the nucleation
rate $J$, the grain number density $n_{\rm gr}$, and the
mass-averaged grain radius $r_{\rm gr}$.
Fig.~\ref {fig:time-tcond_dustval} shows their time evolution
in the top, middle, and bottom panels, respectively.
The left and right columns correspond to
$E_{\rm exp}=10^{48}\,{\rm erg}$ and $10^{51}\,{\rm erg}$,
and the colors indicate the ejecta mass.
For each model, we define $t_{\rm peak}$ as the time
when $J$ reaches its maximum and plot these quantities
against $t-t_{\rm peak}$, so that all nucleation peaks
are aligned at zero.}
\revsec{Table~\ref{tab:Jspeak_summary} lists $t_{\rm peak}$ measured    
from the start of the calculation.
For $E_{\rm exp}=10^{48}\,{\rm erg}$,
$t_{\rm peak}$ decreases from $110$ to $11$ days
as $M_{\rm ej}$ increases from $0.3$ to $30\,M_\odot$.
For $E_{\rm exp}=10^{51}\,{\rm erg}$,
the dependence is monotonically increasing.
The peak time increases from $79$ to $300$ days
between $0.3$ and $30\,M_\odot$.
At fixed ejecta mass, the higher-energy models reach
the nucleation peak later except for $M_{\rm ej} = 0.3 M_\odot$.
}

\revsec{Around the nucleation peak, $n_{\rm gr}$ rises rapidly as new grains form. Subsequent accretion increases $r_{\rm gr}$ toward its final value, producing a step-like evolution.}
At both explosion energies, increasing the ejecta mass increases the final grain radius.
\rev{When fewer grains nucleate, the available carbon monomers are
accreted onto fewer grains, allowing each grain to grow more efficiently.}
\rev{The subsequent grain growth is rapidly suppressed by the decreasing
gas density due to ejecta expansion and by the depletion of carbon
monomers in the gas phase.}

The temporal structure, however, differs between the two explosion energies. For $E_{\rm exp}=10^{48}\,{\rm erg}$, nucleation and rapid grain growth are strongly concentrated around $t_{\rm peak}$. For $E_{\rm exp}=10^{\revsec{51}}\,{\rm erg}$, the nucleation rate can evolve over a broader interval, particularly in the lower-mass models. 
\revsec{Grain growth by gas accretion follows and is completed immediately after the formation of seed nuclei (characterized by $J$), resulting in $r_{\rm gr}$ tracks that resemble a step function. }

\rev{
According to
Eq.~(\ref{eq:grain_growth_rate}), the higher key-species number density gives a larger grain growth rate around $t_{\rm peak}$, resulting
in a larger final grain size.}
The properties of the models with $E_{\rm exp}=10^{48}\,{\rm erg}$ and $10^{51}\,{\rm erg}$ are summarized in Table~\ref{tab:Jspeak_summary}. 
At the same ejecta mass, the higher-energy models generally reach the nucleation peak at lower densities and produce smaller graphite grains. For example, at $M_{\rm ej}=10M_\odot$, increasing the explosion energy from $10^{48}\,{\rm erg}$ to $10^{\revsec{51}}\,{\rm erg}$ reduces the density at the nucleation peak from $3.6\times10^{-6}$ to $1.3\times10^{-14}\,{\rm g\,cm^{-3}}$, and the final grain size from $3.5\times10^3$ to $2.5\times10^{-3}\,{\rm \mu m}$.

\revsec{We next examine the final grain size over a wider range of $E_{\rm exp}$, $M_{\rm ej}$, and $R_0$ in Ni-free adiabatic models.}
Fig.~\ref {fig:Eex_Mej_R0_minipage} shows the final mass-averaged
graphite grain radius on the $E_{\rm exp}$--$M_{\rm ej}$ plane.
Throughout these calculations, we fix $f_{\rm int}=0.9$ and
$X'_{\rm C}=3\times10^{-3}$.
\revsec{The contours at
$r_{\rm gr}=0.01$, $1$, and $100\,{\rm \mu m}$
run approximately from the lower left to the upper right
as straight lines in the logarithmic
$E_{\rm exp}$--$M_{\rm ej}$ plane.}

\revsec{At each fixed $R_0$, the final grain size increases
with increasing $M_{\rm ej}$ and decreasing $E_{\rm exp}$.
The contours of constant grain size therefore shift toward
larger $M_{\rm ej}$ as $E_{\rm exp}$ increases.
In particular, models above the
$r_{\rm gr}=1\,{\rm \mu m}$ contour produce micron-sized
or larger grains, showing that higher explosion energies
require larger ejecta masses to reach this size.}

\revsec{The initial radius also strongly affects the final
grain size.
As $R_0$ increases, the region producing micron-sized grains
contracts toward lower $E_{\rm exp}$ and larger $M_{\rm ej}$.
For example, at $E_{\rm exp}=10^{49}\,{\rm erg}$ and
$M_{\rm ej}=10\,M_\odot$, the final grain size exceeds
$1\,{\rm \mu m}$ for $R_0=10^{12}\,{\rm cm}$ but falls
below this value for $R_0=10^{13}\,{\rm cm}$.
For $R_0=10^{14}\,{\rm cm}$, no models reach
$1\,{\rm \mu m}$ within the parameter range shown in
Fig.~\ref {fig:Eex_Mej_R0_minipage}.}

\subsection{\revsec{Analytical estimates of the grain size}}
\label{subsec:carbon_density_cooling_timescale}

Fig.~\ref {fig:time-tcond_dustval} shows that graphite grains grow mainly during a short interval around the peak of the steady-state nucleation rate. 
\revsec{This suggests that, despite the different evolutionary histories,
the final grain size may be characterized by the local thermodynamic
conditions during the graphite-growth phase.
We therefore seek a simple quantity that connects these local conditions
to the final grain size across different ejecta models.}

\revsec{Guided by the peak nucleation temperatures in
Table~\ref{tab:Jspeak_summary},
we adopt $T=2000\,{\rm K}$ as a representative temperature of the graphite-growth phase. Evaluating Eq.~(\ref{eq:grain_growth_rate}) at this temperature
with the undepleted carbon number density
$X'_{\rm C}\rho_{2000}/m_{1{\rm C}}$, we define
the reference growth rate as}
\begin{align}
\revsec{
    \left. \frac{dr}{dt}\right|_{\rm ref}
    \equiv
    \frac{\alpha_{\rm C}\Omega_{1{\rm C}}v_{{\rm th},2000}}
         {m_{1{\rm C}}}
    X'_{\rm C}\rho_{2000}.
}
\end{align}
\revsec{The characteristic increase in grain radius
is then approximated as
$\Delta r\simeq (dr/dt|_{\rm ref})\Delta t_{\rm eff}$.
We parameterize the effective growth duration as}
\begin{align}   
\revsec{
    \Delta t_{\rm eff}
    = f_{\rm grow}\tau_{{\rm cool},2000}.
}
\end{align}
\revsec{
Here, $\tau_{{\rm cool}, 2000} \equiv |d\ln T/dt|^{-1}_{\rm 2000K}$ is the cooling timescale evaluated when the gas temperature reaches $2000\,{\rm K}$. The dimensionless factor $f_{\rm grow}$ accounts for the finite growth interval, monomer depletion, and changes in the growth rate relative to the reference value.
\revsec{A more detailed derivation is given in Appendix \ref{App:collision_exposure}.}
Retaining $a_{1{\rm C}}$ as the monomer-size baseline,
we obtain the estimate}
\begin{align}
r_{{\rm gr},{\rm final}}
\simeq
a_{1{\rm C}}
+
f_{\rm grow}
\frac{\alpha_{\rm C}\Omega_{1{\rm C}}v_{{\rm th},2000}}
{m_{1{\rm C}}}
X'_{\rm C}\rho_{2000}\tau_{{\rm cool},2000}.
\label{eq:rgr_final_estimate}
\end{align}

\begin{figure}
    \includegraphics[width=1.\linewidth]{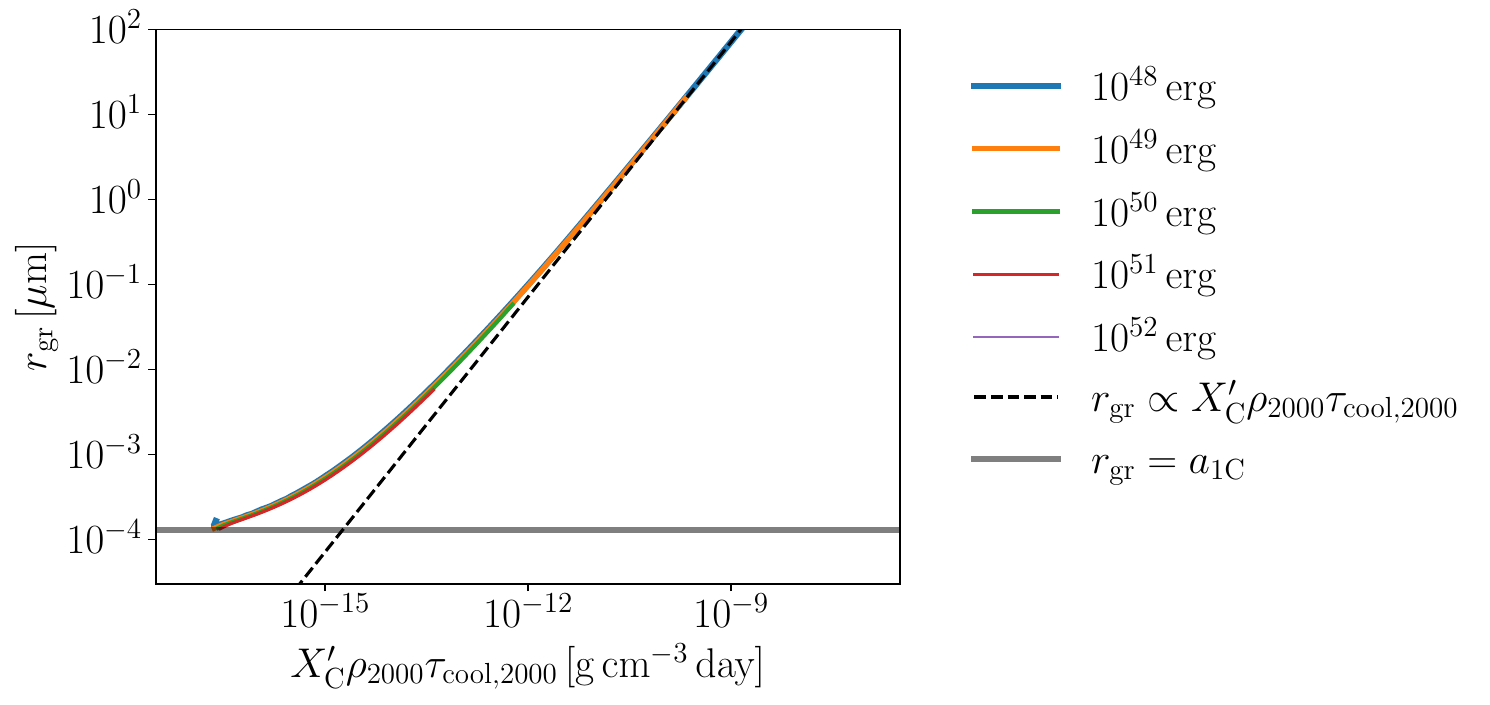}
    \caption{
    Final mass-averaged graphite grain radius as a function of
    $X'_{\rm C}\rho_{2000}\tau_{\rm cool,2000}$ for Ni-free adiabatic models with
    $R_0=10^{13}\,{\rm cm}$.
    Here, $\rho_{2000}$ and $\tau_{\rm cool,2000}$ are the gas density and cooling timescale evaluated when the gas temperature reaches
    $T=2000\,{\rm K}$, and $X'_{\rm C}$ is the effective carbon abundance available for graphite formation.
    Different colored curves correspond to different explosion energies.
    The black dashed line shows a slope-one guide corresponding to
    $r_{\rm gr}\propto X'_{\rm C}\rho_{2000}\tau_{\rm cool,2000}$.
    The gray horizontal line indicates the monomer radius.
    The same collapse is obtained for models with $R_0=10^{11}$--$10^{14}\,{\rm cm}$.
    } 
    \label{fig:universal_rel_size}
\end{figure}

\revsec{Eq.~(\ref{eq:rgr_final_estimate}) suggests $X'_{\rm C}\rho_{2000}\tau_{\rm cool,2000}$ as a useful quantity for characterizing grain growth. However, $f_{\rm grow}$ can vary between models. We therefore examine numerically whether models with similar values of this product yield similar final grain sizes.}
Fig.~\ref {fig:universal_rel_size} shows this comparison for models with $R_0=10^{13}\,{\rm cm}$.
Although the explosion energy is varied, the results approximately collapse onto a single relation. 
We have confirmed that the same collapse is obtained for other values of $R_0$ in the range $10^{11}$--$10^{14}\,{\rm cm}$.
\revsec{The numerical results thus reveal an approximately single-valued relation between $X'_{\rm C}\rho_{2000}\tau_{\rm cool,2000}$ and the final grain size over the explored range of Ni-free adiabatic models. Different combinations of explosion parameters that give similar values of this product yield similar final grain sizes.}

\revsec{For sufficiently large values of
$X'_{\rm C}\rho_{2000}\tau_{\rm cool,2000}$ $(\gtrsim 10^{-12}\, {\rm g\, cm^{-3}\,day})$, }
the numerical results are broadly consistent with a linear relation between $r_{\rm gr,final}$ and $X'_{\rm C}\rho_{2000}\tau_{\rm cool,2000}$. 
At small \rev{$X'_{\rm C}\rho_{2000}\tau_{\rm cool,2000}$} \revsec{($\sim 10^{-16}\, {\rm g\, cm^{-3}\,day}$)},
the grain size approaches the monomer size, indicating that substantial growth does not occur.
In the present calculations, a graphite grain size of
$r_{\rm gr,final}\simeq1\,{\rm \mu m}$ corresponds approximately to
\begin{align}
X'_{\rm C}\rho_{2000}\tau_{\rm cool,2000}
\simeq
1.2\times10^{-11}\,{\rm g\,cm^{-3}\,day}.
\end{align}
This value is an empirical threshold obtained from the present graphite-growth calculations rather than a first-principles analytic boundary.
Because this relation is expressed entirely in terms of local quantities during the graphite-growth phase, it does not explicitly depend on the equation of state.
The equation of state enters when $\rho_{2000}$ and $\tau_{\rm cool,2000}$ are mapped onto the macroscopic ejecta parameters, such as $E_{\rm exp}$, $M_{\rm ej}$, and $R_0$.
We examine this mapping in the next subsection.

\subsection{\rev{Dependence of graphite grain size on supernova parameters in Ni-free ejecta}}
\label{subsec:Ni-free-search}

\revsec{
We now interpret the parameter dependence shown in Fig.~\ref {fig:Eex_Mej_R0_minipage} using the grain-growth estimate developed in Section~\ref{subsec:carbon_density_cooling_timescale}.
Specifically, we derive how $X'_{\rm C}\rho_{2000}\tau_{\rm cool,2000}$ depends on $E_{\rm exp}$, $M_{\rm ej}$, and $R_0$ in the matter-dominated adiabatic limit.}
\rev{
Mapping this quantity onto $E_{\rm exp}$, $M_{\rm ej}$, and $R_0$, however, depends on the thermal evolution of the ejecta.}
For adiabatic expansion with an approximately constant adiabatic index, the temperature and density satisfy $T\propto\rho^{\gamma-1}$. Thus, in the radiation- and matter-dominated limits,
\begin{align}
T \propto
\begin{cases}
    \rho^{1/3} \qquad {\rm (radiation\ dominated)},\\
    \rho^{2/3} \qquad {\rm (matter\ dominated)}.
\end{cases}
\end{align}

Since graphite growth occurs around $T\simeq2000\,{\rm K}$, the product $X'_{\rm C}\rho_{2000}\tau_{\rm cool,2000}$ at $2000\, {\rm K}$ is important for grain growth. For a given decrease in temperature, radiation-dominated ejecta experience stronger density dilution than matter-dominated ejecta and are therefore less favorable for forming large grains.
\rev{Here, we focus on the matter to matter adiabatic case, for which a simple analytic scaling can be derived.}

For the matter to matter case, as shown in Appendix \ref{App:scaling_of_mat_dom}, the density and cooling timescale at $T=2000\,{\rm K}$ scale as
$\rho_{2000}\propto R_0^{-3}M_{\rm ej}^{5/2}E_{\rm exp}^{-3/2}$
and
$\tau_{\rm cool,2000}\propto R_0$,
respectively. 
\rev{Eq.~(\ref{eq:rgr_final_estimate}) identifies $X'_{\rm C}\rho_{2000}\tau_{\rm cool,2000}$ as the local combination that controls grain growth 
(a physical motivation for this combination, based on the timescale ratio of \citet{Yamamoto1977PThPh}, is given in Section~\ref{subsec:YH_comparison}). 
Here, restricting ourselves to the efficient-growth branch
in the matter-dominated adiabatic limit, we relate this local quantity
to the macroscopic ejecta parameters $E_{\rm exp}$, $M_{\rm ej}$, and $R_0$.}
Therefore,
\rev{$X'_{\rm C}\rho_{2000}\tau_{\rm cool,2000}$ scales as}     
\begin{align}
X'_{\rm C}\rho_{2000}\tau_{\rm cool,2000}
\propto
X'_{\rm C}
E_{\rm exp}^{-3/2}
M_{\rm ej}^{5/2}
R_0^{-2}.
\label{eq:local_growth_scaling_mat}
\end{align}

\rev{This scaling reproduces the numerical trend that grain growth is
favored by lower $E_{\rm exp}$, larger $M_{\rm ej}$, and smaller $R_0$.
The analytic argument determines the parameter dependence but not its
normalization. Using the numerical normalization of the efficient-growth
branch in Fig.~\ref {fig:universal_rel_size}, we obtain}
\begin{align}
r_{\rm gr,final}
&\simeq
615\,{\rm \mu m}
\left(\frac{E_{\rm exp}}{10^{48}\,{\rm erg}}\right)^{-3/2}
\left(\frac{M_{\rm ej}}{5\,M_\odot}\right)^{5/2}
\notag\\
&\times
\left(\frac{R_0}{10^{12}\,{\rm cm}}\right)^{-2}
\left(\frac{X'_{\rm C}}{3\times10^{-3}}\right).
\label{eq:rgr_mat_scaling}
\end{align}
\rev{Eq.~(\ref{eq:rgr_mat_scaling})}
should be interpreted as a numerical fit to the efficient-growth branch in the matter-dominated adiabatic limit, not as the radius obtained by allowing grains to grow over the full local cooling time.

For reference, setting $r_{\rm gr,final}=1\,\mu{\rm m}$
\rev{in Eq.~(\ref{eq:rgr_mat_scaling})} gives the formal micron-size contour, 
\begin{align}
M_{\rm ej,1\mu m}^{\rm mat}
&\simeq
0.38M_\odot
\left(\frac{E_{\rm exp}}{10^{48}\,{\rm erg}}\right)^{3/5}
\left(\frac{R_0}{10^{12}\,{\rm cm}}\right)^{4/5}
\notag\\
&\quad\times
\left(\frac{X'_{\rm C}}{3\times10^{-3}}\right)^{-2/5}.
\label{eq:micron_contour_mat}
\end{align}
\rev{The formal contour in Eq.~(\ref{eq:micron_contour_mat}) is meaningful only}
 where the ejecta is actually matter dominated during the cooling from the initial state to the graphite-condensation phase.

A useful self-consistency condition for the matter-dominated branch is obtained from the ratio of the initial radiation and gas internal energies,
\begin{align}
\eta_0
:=
\frac{\varepsilon_{\rm rad,0}}{\varepsilon_{\rm \rev{mat},0}}
&\simeq
1.0\times10^{-2}
\left(\frac{E_{\rm exp}}{10^{48}\,{\rm erg}}\right)^3
\left(\frac{M_{\rm ej}}{5\,M_\odot}\right)^{-4}
\notag \\
&\times
\left(\frac{R_0}{10^{12}\,{\rm cm}}\right)^3
\left(\frac{f_{\rm int}}{0.9}\right)^3 .
\label{eq:eta0}
\end{align}
The matter-dominated scaling is self-consistent only for $\eta_0\lesssim1$, or equivalently
\begin{align}
M_{\rm ej}
\gtrsim
1.6\,M_\odot
\left(\frac{E_{\rm exp}}{10^{48}\,{\rm erg}}\right)^{3/4}
\left(\frac{R_0}{10^{12}\,{\rm cm}}\right)^{3/4}
\left(\frac{f_{\rm int}}{0.9}\right)^{3/4}.
\label{eq:matter_dom_condition}
\end{align}

\rev{Comparing the four panels at fixed $E_{\rm exp}$ and $M_{\rm ej}$
shows that the final grain size systematically decreases as $R_0$
increases. In particular, the $1\,\mu{\rm m}$ contour shifts toward
lower $E_{\rm exp}$ and higher $M_{\rm ej}$ with increasing $R_0$.}
\rev{Thus, the analytic scaling provides a physical interpretation of the numerical trends in Fig.~\ref {fig:Eex_Mej_R0_minipage}, while the full one-zone calculations determine the grain sizes outside the self-consistent matter-dominated regime.}
These results show that, in the Ni-free adiabatic limit, low-energy
and massive ejecta provide the most favorable hydrodynamical conditions
for forming micron-sized graphite grains.
We next examine how this picture is modified by radioactive heating
and radiative cooling.
\subsection{\revsec{The effect of Ni heating}} 
\label{subsec:Ni_heating}
\begin{figure*}
    \begin{center}
    \includegraphics[width=0.9\linewidth]{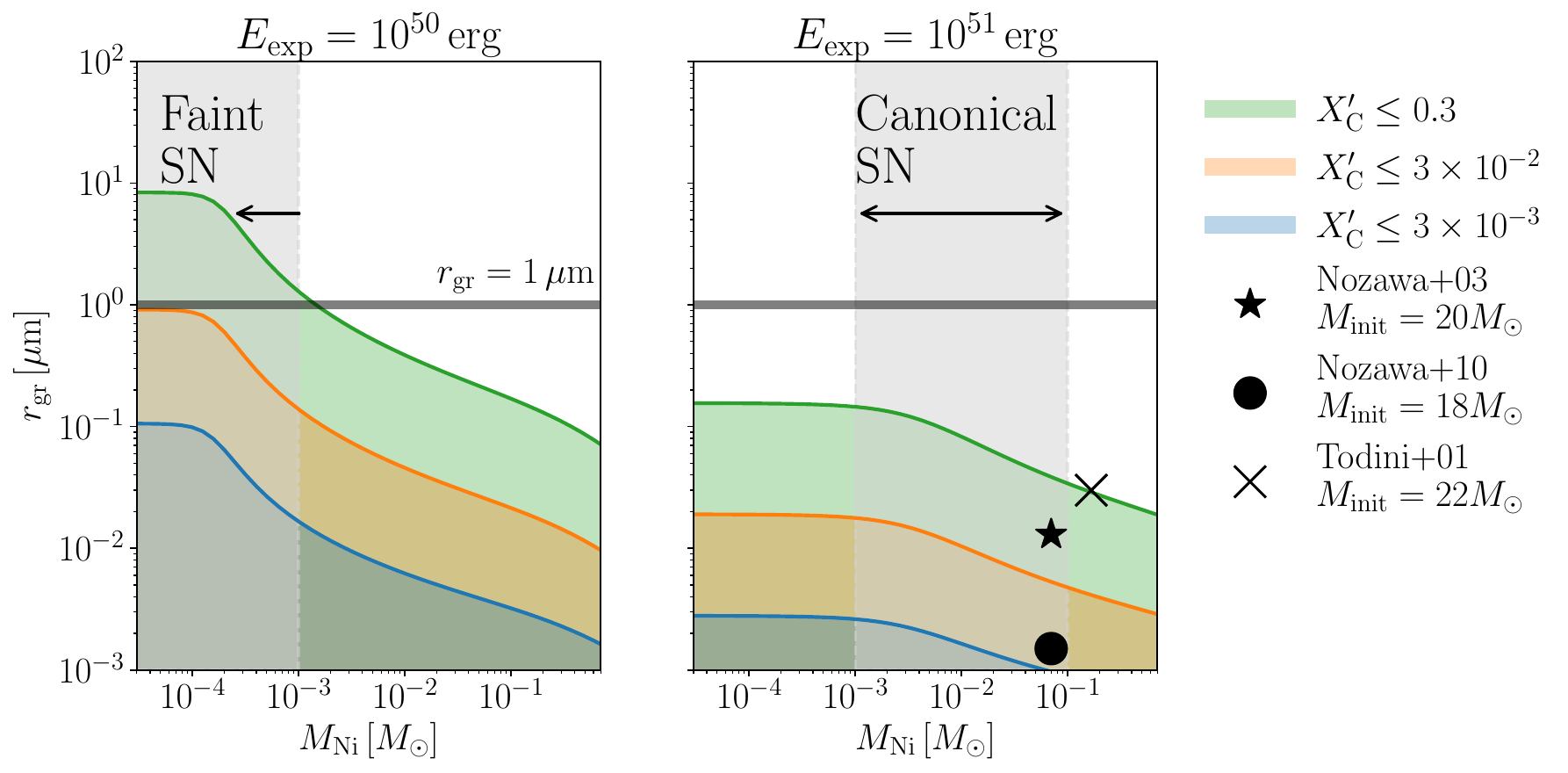}
    \end{center}
    \caption{The graphite grain sizes ($r_{\rm gr}$) as a function of the ejected $^{56}\mathrm{Ni}$ mass ($M_{\rm Ni}$). The left and right panels show models with $E_{\rm exp}=10^{50}\,{\rm erg}$ and $E_{\rm exp}=10^{51}\,{\rm erg}$, respectively. 
    \rev{The green, orange, and blue curves correspond to $X'_{\rm C}=3\times10^{-1}$, $3\times10^{-2}$, and $3\times10^{-3}$, respectively, while the corresponding shaded regions show the ranges $X'_{\rm C}\leq3\times10^{-1}$, $X'_{\rm C}\leq3\times10^{-2}$, and $X'_{\rm C}\leq3\times10^{-3}$.}
    The gray shaded regions are labeled ``Faint SN'' in the left panel and ``Canonical SN'' in the right panel. The horizontal gray line marks $r_{\rm gr}=1\,{\rm \mu m}$.
    \rev{The symbols show representative graphite grain sizes from previous supernova dust-formation calculations by \citet{Nozawa2003ApJ}, \citet{Nozawa2010ApJ}, and \citet{Todini2001MNRAS}, for the progenitor masses indicated in the legend.}}
    \label{fig:MNi_rgr_R01e12}
\end{figure*}

\revsec{In Section~\ref{subsec:Ni-free-search}, we considered Ni-free adiabatic models with a fixed effective carbon abundance of $X'_{\rm C}=3\times10^{-3}$. 
We now include radioactive heating and radiative cooling, and examine how the grain size depends on $E_{\rm exp}$, $M_{\rm Ni}$, and $X'_{\rm C}$, while fixing $R_0=10^{12}\,{\rm cm}$, $M_{\rm ej}=10\,M_\odot$, and $f_{\rm int}=0.9$.}
Figure \ref{fig:MNi_rgr_R01e12} shows the representative graphite grain size as a function of $M_{\rm Ni}$. The left and right panels show models with $E_{\rm exp}=10^{50}\,{\rm erg}$ and $10^{51}\,{\rm erg}$, respectively. The blue, orange, and green curves correspond to $X'_{\rm C}=3\times10^{-3}$, $3\times10^{-2}$, and $3\times10^{-1}$, while the shaded regions indicate the ranges of $X'_{\rm C}$ adopted for the three illustrative composition cases.
\rev{These three values are chosen as illustrative benchmarks spanning the carbon abundances from the H-rich envelope to He-rich and carbon--oxygen-rich layers of massive-star progenitors \citep{Umeda2012PTEP, Asplund2021A&A}.}
\rev{The gray shaded regions indicate the faint-SN ($M_{\rm Ni}<10^{-3}\,M_\odot$) and canonical-SN ($10^{-3}\leq M_{\rm Ni}/M_\odot\leq10^{-1}$) regimes adopted here, motivated by the $^{56}$Ni mass ranges discussed in Section~\ref{sec:Intro}.}
\revsec{The symbols show grain sizes from \citet{Nozawa2003ApJ} and \citet{Nozawa2010ApJ} for progenitor masses of $20\,M_\odot$ and $18\,M_\odot$, respectively. 
The cross shows the result for the $22\,M_\odot$ model of \citet{Todini2001MNRAS}.}

\rev{Fig.~\ref {fig:MNi_rgr_R01e12} shows three systematic trends.
At fixed $E_{\rm exp}$ and $M_{\rm Ni}$, the final grain size increases
with increasing $X'_{\rm C}$. At fixed $M_{\rm Ni}$ and $X'_{\rm C}$,
lower explosion energy produces larger grains. For both explosion
energies, increasing $M_{\rm Ni}$ suppresses grain growth, while the
dependence on $M_{\rm Ni}$ becomes weak toward the low-$M_{\rm Ni}$ regime.}
\rev{At sufficiently small $M_{\rm Ni}$, radioactive heating becomes
subdominant, and the thermal evolution is governed mainly by adiabatic
expansion and radiative cooling. In the limiting case $M_{\rm Ni}=0$,
the resulting grain sizes approach those obtained in the Ni-free
adiabatic calculations discussed in the previous subsection.}

\revsec{
We can understand these features as follows.
In Eq.~(\ref{eq:rgr_final_estimate}), the final grain size is
set by the product $\rho_{2000}\tau_{{\rm cool},2000}$.
Radioactive heating lengthens $\tau_{{\rm cool},2000}$, but it also
delays condensation while the ejecta continue to expand, lowering
$\rho_{2000}$ by orders of magnitude (Fig.~\ref{fig:rho-T_plane}).
The latter dominates, and the resulting decrease in the product
reduces $r_{{\rm gr},{\rm final}}$.
}

\rev{For $E_{\rm exp}=10^{51}\,{\rm erg}$, micron-sized graphite does
not form over the explored range of $X'_{\rm C}$.
Within the canonical-SN regime, even the largest adopted carbon abundance
gives $r_{\rm gr}\lesssim0.1$--$0.2\,\mu{\rm m}$, with progressively
smaller grains toward lower $X'_{\rm C}$.}
\rev{By contrast, in the faint-SN regime with
$E_{\rm exp}=10^{50}\,{\rm erg}$, graphite grains with radii of order
$1$--$10\,\mu{\rm m}$ can form for
$X'_{\rm C}\sim3\times10^{-2}$--$3\times10^{-1}$,
representative of the He layer and carbon--oxygen core.}

\section{Discussion}
\label{sec:discussion}

\subsection{\revsec{Implications for the origin of presolar graphite}}
\revsec{Our results identify low-energy, nickel-poor explosions
as possible sources of micron-sized presolar graphite.
In the models shown in Fig.~\ref {fig:MNi_rgr_R01e12},
with $M_{\rm ej}=10\,M_\odot$, $R_0=10^{12}\,{\rm cm}$,
and $f_{\rm int}=0.9$, explosions with
$E_{\rm exp}=10^{50}\,{\rm erg}$ and
$M_{\rm Ni}\lesssim10^{-3}\,M_\odot$ can produce
graphite grains with sizes of order $1$--$10\,\mu{\rm m}$
for $X'_{\rm C}\sim0.03$--$0.3$.
These sizes are comparable to those of large presolar
graphite grains identified as supernova condensates
\citep{Amari2014GeCoA}.
Thus, the discrepancy between measured presolar grain sizes
and previous calculations may partly reflect the range
of explosion parameters considered.
}

\revsec{
\revfourth{Observationally inferred explosion parameters indicate that such low-energy, nickel-poor explosions are not rare.}
\citet{Martinez2022A&A} obtained median explosion energies of $6.1\times10^{50}\,{\rm erg}$ for their gold sample of type II supernovae and $6.3\times10^{50}\,{\rm erg}$ for their full sample, below the commonly adopted $10^{51}\,{\rm erg}$. Nickel masses also span a broad distribution. In the type II supernova comparison sample used by \citet{Meza2020A&A}, more than $25\%$ have $M_{\rm Ni}<0.015\,M_\odot$. Furthermore, \citet{Muller2017ApJ} found a positive correlation between explosion energy and nickel mass. Lower-energy explosions therefore tend to combine two properties that favor grain growth in our models. These observations motivate examining the low-energy, low-nickel part of the population as a source of large graphite grains. However, they do not establish the fraction of supernovae that satisfy the micron-grain formation conditions identified here.}

\revfourth{The ejecta mass is also constrained observationally. In the sample of
\citet{Martinez2022A&A}, $M_{\rm ej}$ is concentrated near $8$--$10\,M_\odot$,
with a median of $8.4\,M_\odot$ for the gold sample, and is only weakly correlated
with the explosion energy (Pearson coefficient $0.36\pm0.11$; \cite{Martinez2022bA&A}).
The observed type II supernovae therefore occupy a band of $M_{\rm ej}\sim10\,M_\odot$
and $E_{\rm exp}\gtrsim10^{50}\,{\rm erg}$ in Fig.~\ref{fig:Eex_Mej_R0_minipage};
the upper-left region of the figure, $E_{\rm exp}\sim10^{48}\,{\rm erg}$ with
$M_{\rm ej}\sim10\,M_\odot$, where our models yield the largest grains, has no
observed supernova counterpart.}

\revsec{The contribution of these grains to planetary systems
also depends on their survival and transport.
Grain-size-dependent shock processing can modify the
population that survives after formation
\citep{Nozawa2006ApJ, Nozawa2007ApJ}.
Grain size also affects injection into protoplanetary disks.
The \revfourth{short-lived radionuclide} enrichment model of \citet{Sawada2024ApJ} includes direct injection by supernova dust grains with radii $\gtrsim1\,\mu{\rm m}$, which enter the disk more efficiently than small grains. Our calculations show that graphite can reach this size under favorable explosion conditions. They therefore motivate accounting for the dependence of grain formation on explosion parameters when modeling solid delivery to disks. Quantifying radionuclide delivery additionally requires the incorporation of the relevant nuclides into dust, together with grain survival and transport.}

\revsec{A further test is whether the same ejecta conditions
can reproduce both the grain sizes and isotopic compositions.
Measurements of C, O, Si, and Ti isotopes in individual
graphite grains provide constraints on their source material
and mixing between ejecta layers \citep{Travaglio1999ApJ, Jadhav2013GeCoA}.
In particular, \citet{Schofield2022MNRAS} found that their low- and
intermediate-energy models without a C/Si zone require
mixing with oxygen-rich ejecta to explain the
$^{28}$Si and $^{44}$Ti signatures.
Testing the low-energy origin proposed here therefore
requires coupling grain formation to nucleosynthesis
and ejecta mixing.}

\subsection{\revsec{Comparison with Yamamoto and Hasegawa (1977)}}
\label{subsec:YH_comparison}

\revfourth{The quantity $X'_{\rm C}\rho_{2000}\tau_{\rm cool,2000}$ examined in this work revisits the timescale ratio $\Lambda_{\rm on}$ introduced by \citet{Yamamoto1977PThPh} and applied to supernova ejecta by \citet{Nozawa2013ApJ}.}
\revsec{\citet{Yamamoto1977PThPh} characterized grain formation using the ratio
of the supersaturation timescale to the monomer collision timescale.
Following \citet{Nozawa2013ApJ}, we consider this ratio at the onset of dust
formation, $t_{\rm on}$, before appreciable monomer depletion.
For graphite, the undepleted carbon number density is
$\tilde{c}_{1{\rm C}}=X'_{\rm C}\rho/m_{1{\rm C}}$,
and the timescale ratio can be written as}
\begin{align}
\revsec{
\Lambda_{{\rm on},{\rm C}}
\equiv
\frac{\tau_{{\rm sat},{\rm C}}(t_{\rm on})}
     {\tau_{{\rm coll},{\rm C}}(t_{\rm on})}
=
\left.
\frac{4\pi\alpha_{\rm C}a_{1{\rm C}}^2}{m_{1{\rm C}}}
\frac{v_{{\rm th},{\rm C}}}
     {|d\ln S_{\rm C}/d\ln T|}
X'_{\rm C}\rho\tau_{\rm cool}
\right|_{t=t_{\rm on}} .
}
\label{eq:Lambda_on_C}
\end{align}
\revsec{Here,
$\tau_{{\rm sat},{\rm C}}\equiv|d\ln S_{\rm C}/dt|^{-1}$,
$\tau_{{\rm coll},{\rm C}}
=(4\pi\alpha_{\rm C}a_{1{\rm C}}^2
\tilde{c}_{1{\rm C}}v_{{\rm th},{\rm C}})^{-1}$,
and $\tau_{\rm cool}\equiv|d\ln T/dt|^{-1}$.}

\revfourth{Before appreciable depletion, $|d\ln S_{\rm C}/d\ln T|\simeq A/T\simeq40$
at $T=2000\,{\rm K}$ for adiabatic cooling (Appendix~\ref{App:collision_exposure}),
so that $\tau_{{\rm sat},{\rm C}}\simeq(T/A)\tau_{\rm cool}$, as in the
approximation used by \citet{Nozawa2013ApJ}, and Eq.~(\ref{eq:Lambda_on_C})
reduces to $\Lambda_{{\rm on},{\rm C}}\propto X'_{\rm C}\rho\tau_{\rm cool}$.
This provides the physical motivation for our use of
$X'_{\rm C}\rho_{2000}\tau_{{\rm cool},2000}$, with $2000\,{\rm K}$ adopted as a
common reference temperature, whereas $\Lambda_{{\rm on},{\rm C}}$ is evaluated
at the actual onset of dust formation.
Expressed in this way, the timescale ratio is no longer tied to a particular
ejecta model, 
but can be mapped directly onto the macroscopic explosion parameters, as was done in
Section~\ref{subsec:Ni-free-search}
to obtain the scaling of Eq.~(\ref{eq:local_growth_scaling_mat}) with
$E_{\rm exp}$, $M_{\rm ej}$, and $R_0$.}

\subsection{\revsec{Caveat}}

\revsec{
The present one-zone calculations identify conditions favorable
for graphite formation, but do not resolve the density and
composition structure, clumping, or mixing within the ejecta.
The adopted carbon abundances therefore represent illustrative
compositions rather than individual ejecta layers with distinct
thermal histories.
In particular, the absence of micron-sized grains in our
canonical-energy models does not exclude their formation
in locally dense, carbon-rich regions.
Conversely, the favorable conditions found at low explosion
energies require that carbon-rich material be ejected.
}

\revfourth{Whether such material escapes the progenitor depends on fallback and on mixing within the ejecta. In one-dimensional models without mixing, carbon--oxygen core material is ejected only when $E_{\rm exp}$ exceeds the binding energy of the hydrogen envelope \citep{Shinoda2026MNRAS}. Even above this threshold, the amount of helium-layer or carbon--oxygen core material that escapes depends on mixing, which has been parameterized in one-dimensional models as the mixing-fallback framework \citep{Umeda2002ApJ, Tominaga2007ApJ}; a quantitative treatment requires multi-dimensional calculations \citep{Tominaga2009ApJ, Chan2020MNRAS}.}

\revsec{
The available carbon abundance $X'_{\rm C}$ is prescribed,
without explicitly following the formation and destruction
of CO or other molecules.
The effects of radioactive heating are thus evaluated
at a specified carbon availability, while radioactive decay
may also affect the chemistry that determines this availability.
The thermal evolution is likewise approximate, with a fixed
gas equation of state, constant opacities, and simplified
radiative cooling and radioactive energy deposition.
These assumptions introduce uncertainties in the temperature
history and hence in the density and cooling timescale
during grain formation.
}

\revsec{
The dust calculation adopts steady-state classical nucleation
theory, bulk material properties for small clusters, a unit
sticking probability, and equal gas and condensed-phase
temperatures.
The sensitivity of the predicted grain sizes to these assumptions,
including departures from steady-state nucleation, remains
to be assessed.
Finally, we calculate grain sizes at formation and do not
follow subsequent processing by reverse shocks or destruction
in the interstellar medium.
Connecting these results to presolar graphite therefore
requires both a viable formation environment and the survival
of the resulting grains.
}

\section{Conclusions}
\label{sec:conclusion}
\revsec{
We investigated graphite formation in supernova ejecta
by combining a one-zone model of the ejecta evolution
with calculations of nucleation and grain growth.
We examined how the explosion parameters, radioactive heating,
and available carbon abundance affect the formation
of micron-sized graphite.
}

\revsec{
For the Ni-free adiabatic models, the final grain radius
approximately follows a single relation with
$X'_{\rm C}\rho_{2000}\tau_{\rm cool,2000}$
over the explored parameter range.
This numerical relation connects the final grain size
to the carbon supply and cooling timescale during
the grain-growth phase.
In the efficient-growth regime, the grain radius
is approximately proportional to this product.
For ejecta that remain matter dominated during adiabatic cooling,
the product scales as
$X'_{\rm C}E_{\rm exp}^{-3/2}M_{\rm ej}^{5/2}R_0^{-2}$,
linking efficient grain growth to lower explosion energies,
larger ejecta masses, and smaller initial radius
within this regime.
}


\revsec{
Our results identify low-energy explosions with small
ejected $^{56}{\rm Ni}$ masses as promising environments
for forming large graphite grains, provided that
sufficient carbon-rich material is ejected.
The conditions identified here provide a basis
for testing a faint-supernova origin of large presolar graphite
with models that resolve the ejecta structure and chemistry
and follow the subsequent survival of the grains.
}



\begin{ack}
K.S. thanks Shinichi Shirono and Satoshi Okuzumi for fruitful discussions.
\end{ack}

\section*{Funding}
K.S. acknowledges support from JST SPRING, Grant Number JPMJSP2108. 
This work was also supported by JSPS KAKENHI Grant Numbers JP25K07351 (R.T.), JP26KJ0211 (T.T.), JP24H02245, JP24K00668, and JP25K01035 (Y.S.).

\section*{Data availability} 
 The data underlying this article will be shared on reasonable request to the corresponding author.

\appendix 

\section{Analytic scaling in the matter-dominated adiabatic limit}
\label{App:scaling_of_mat_dom}

We derive the scalings used in Section~\ref{subsec:Ni-free-search}
for Ni-free, matter-dominated adiabatic ejecta with $\gamma=5/3$.
For the analytic estimate we adopt a coasting expansion,
\begin{align}
R(t)&=R_0(1+t/t_0), & t_0&=R_0/v,\label{eq:app_coasting}\\
v&=\left(\frac{2E_{\rm exp}}{M_{\rm ej}}\right)^{1/2}.
\end{align}
The gas-only initial temperature and density are
\begin{align}
T_0&=\frac{(\gamma-1)m_p}{k_B}
\frac{f_{\rm int}E_{\rm exp}}{M_{\rm ej}},\\
\rho_0&=\frac{3M_{\rm ej}}{4\pi R_0^3}.
\end{align}
Mass conservation and $T\propto\rho^{\gamma-1}$ give
$\rho(t)/\rho_0=(1+t/t_0)^{-3},{T(t)}/{T_0}=(1+t/t_0)^{-2}.$
At any reference temperature $T_{\rm r}\leq T_0$, these relations imply
\begin{align}
1+\frac{t_{\rm r}}{t_0}&=\left(\frac{T_0}{T_{\rm r}}\right)^{1/2},
& t_{\rm r}&=t_0\left[\left(\frac{T_0}{T_{\rm r}}\right)^{1/2}-1\right].
\end{align}
Differentiating the temperature history gives
\begin{align}
\frac{d\ln T}{dt}&=-\frac{2}{t+t_0},
&\tau_{\rm cool}(t)&\equiv\left|\frac{d\ln T}{dt}\right|^{-1}
=\frac{t+t_0}{2}.
\end{align}
Consequently,
\begin{align}
\tau_{\rm cool,r}
&=\frac{R_0}{2v}\left(\frac{T_0}{T_{\rm r}}\right)^{1/2}\notag\\
&=7.80\,{\rm day}
\left(\frac{R_0}{10^{12}\,{\rm cm}}\right)
\left(\frac{f_{\rm int}}{0.9}\right)^{1/2}
\left(\frac{T_{\rm r}}{2000\,{\rm K}}\right)^{-1/2}.
\label{eq:app_tau_scaled}
\end{align}
The factors $(M_{\rm ej}/E_{\rm exp})^{1/2}$ from $v^{-1}$ and
$(E_{\rm exp}/M_{\rm ej})^{1/2}$ from $T_0^{1/2}$ cancel.
This cancellation assumes fixed $f_{\rm int}$.
The corresponding density is
\begin{align}
\rho_{\rm r}&=\frac{3M_{\rm ej}}{4\pi R_0^3}
\left[\frac{(\gamma-1)m_pf_{\rm int}E_{\rm exp}}
{k_BT_{\rm r}M_{\rm ej}}\right]^{-3/2}\notag\\
&=3.40\times10^{-7}\,{\rm g\,cm^{-3}}
\left(\frac{E_{\rm exp}}{10^{48}\,{\rm erg}}\right)^{-3/2}
\left(\frac{M_{\rm ej}}{5M_\odot}\right)^{5/2}\notag\\
&\quad\times\left(\frac{R_0}{10^{12}\,{\rm cm}}\right)^{-3}
\left(\frac{f_{\rm int}}{0.9}\right)^{-3/2}
\left(\frac{T_{\rm r}}{2000\,{\rm K}}\right)^{3/2}.
\label{eq:app_rho_scaled}
\end{align}
Multiplication of Eqs.~(\ref{eq:app_tau_scaled}) and (\ref{eq:app_rho_scaled}) yields
\begin{align}
X'_{\rm C}\rho_{\rm r}\tau_{\rm cool,r}
&=7.95\times10^{-9}\,{\rm g\,cm^{-3}\,day}
\left(\frac{X'_{\rm C}}{3\times10^{-3}}\right)\notag\\
&\quad\times\left(\frac{E_{\rm exp}}{10^{48}\,{\rm erg}}\right)^{-3/2}
\left(\frac{M_{\rm ej}}{5M_\odot}\right)^{5/2}
\left(\frac{R_0}{10^{12}\,{\rm cm}}\right)^{-2}\notag\\
&\quad\times\left(\frac{f_{\rm int}}{0.9}\right)^{-1}
\left(\frac{T_{\rm r}}{2000\,{\rm K}}\right).
\label{eq:app_exposure_scaled}
\end{align}
For $T_{\rm r}=2000\,{\rm K}$, this is the product used in the main text.
If the effective growth factor and microscopic properties are fixed,
Eq.~(\ref{eq:rgr_final_estimate}) then gives
\begin{align}
r_{\rm gr,final}-a_{1{\rm C}}&\propto
X'_{\rm C}f_{\rm int}^{-1}
E_{\rm exp}^{-3/2}M_{\rm ej}^{5/2}R_0^{-2}.
\end{align}
At fixed $R_0$, $X'_{\rm C}$, and $f_{\rm int}$,
constant-size contours therefore satisfy
$M_{\rm ej}\propto E_{\rm exp}^{3/5}$.
Here the effective growth factor is assumed to be approximately constant.

To test the gas-only approximation, we evaluate the initial ratio of
radiation to gas energy using the gas-only $T_0$,
\begin{align}
\eta_0=\frac{2am_p}{3k_B}\frac{T_0^3}{\rho_0}
\label{eq:app_eta_scaled}
\end{align}
Along the gas adiabat, $\eta(T)=\eta_0(T/T_0)^{3/2}$,
so radiation becomes less important as the ejecta cools.
The gas-dominated limit requires $\eta_0\ll1$; the approximate
transition $\eta_0\lesssim1$ corresponds to
\begin{align}
M_{\rm ej}&\gtrsim1.58M_\odot
\left(\frac{E_{\rm exp}}{10^{48}\,{\rm erg}}\right)^{3/4}
\left(\frac{R_0}{10^{12}\,{\rm cm}}\right)^{3/4}\notag\\
&\quad\times\left(\frac{f_{\rm int}}{0.9}\right)^{3/4}.
\end{align}
Near this transition the full gas-plus-radiation equation of state
is needed for quantitative results.

\section{Effective growth window}
\label{App:collision_exposure}

Here we give a physical interpretation of the product $X'_{\rm C}\rho_{2000}\tau_{\rm cool,2000}$ used in Section~\ref{subsec:carbon_density_cooling_timescale}.
It combines the mass density of carbon available for grain growth with the local cooling timescale at $T=2000\,{\rm K}$.
Let $n_1$ be the number density of carbon key species, $n_{\rm gr}$ the grain number density.
We use the same flux speed as in the main text,
$v_{\rm th}=\sqrt{k_BT/(2\pi m_{1{\rm C}})}$.

If grains grew over the full local cooling time, the size increment would be proportional to $n_{1,2000}v_{{\rm th},2000}\tau_{\rm cool,2000}$, or equivalently to $X'_{\rm C}\rho_{2000}\tau_{\rm cool,2000}$. 
However, this overestimates the actual grain growth 
because stable grains exist and grow only during a short interval around the nucleation burst. Before this burst, the number of stable grains is negligible. After the burst, the rapid increase in the total grain surface area, $4\pi n_{\rm gr} r_{\rm gr}^2$, shortens the depletion timescale and reduces the supersaturation ratio, thereby quenching both further nucleation and growth.

Thus, the effective collision-active duration is not the full cooling time but
\begin{align}
\Delta t_{\rm coll}
=f_{\rm grow}\tau_{\rm cool,2000},
\end{align}
where $f_{\rm grow}\ll1$ represents the fraction of the cooling time during which monomer--grain collisions effectively contribute to grain growth.

The smallness of $f_{\rm grow}$ is expected from the sharp temperature dependence of the steady-state nucleation rate. The nucleation burst opens the growth window, while monomer depletion closes it. If depletion closes the growth window on a comparable timescale,
the effective collision duration can be estimated from the temperature
width over which nucleation becomes efficient,
\begin{align}
f_{\rm grow}
\sim
\frac{\Delta t_{\rm coll}}{\tau_{\rm cool}}
\sim
\Delta\ln T_{\rm nuc}.
\end{align}
Because the nucleation rate depends exponentially on the supersaturation ratio, this temperature width is much smaller than unity in logarithmic temperature.

We now estimate this temperature width from the strongly temperature-sensitive
exponential factor of the steady-state nucleation rate. Ignoring slowly
varying prefactors, we write
\begin{align}
J&\propto\exp\left(-\frac{\mathcal B_{\rm nuc}}{s^2}\right),\\
s&\equiv\ln S,\qquad \mathcal B_{\rm nuc}\equiv\frac{4\mu^3}{27}.
\end{align}
Here $S$ is the supersaturation ratio and $\mu$ is the dimensionless
surface-energy parameter. For graphite,
\begin{align}
\mu&=\frac{4\pi a_{1{\rm C}}^2\zeta_{\rm C}}{k_BT}\notag\\
\simeq10.45
&\left(\frac{a_{1{\rm C}}}{1.281\times10^{-8}\,{\rm cm}}\right)^2\left(\frac{\zeta_{\rm C}}{1400\,{\rm erg\,cm^{-2}}}\right)
\left(\frac{T}{2000\,{\rm K}}\right)^{-1}.
\end{align}

The logarithmic variation over a small temperature interval is
$|\Delta\ln J|\simeq|d\ln J/d\ln T|\,|\Delta\ln T|$.
Using $\mathcal B_{\rm nuc}\propto T^{-3}$ gives
\begin{align}
\frac{d\ln J}{d\ln T}
&\simeq\frac{3\mathcal B_{\rm nuc}}{s^2}
+\frac{2\mathcal B_{\rm nuc}}{s^3}\frac{ds}{d\ln T}.
\label{eq:app_J_derivative}
\end{align}
Around the nucleation epoch, $s$ changes rapidly because it contains
both the equilibrium and carbon partial-pressure terms.
Before depletion on a matter-dominated adiabat,
$q\equiv-ds/d\ln T\simeq A/T-5/2\simeq40.7$ at $2000\,{\rm K}$.
At $(\mu,s,q)=(10,4,40)$, the slope magnitude is approximately 157.
The corresponding local one-e-fold temperature interval is
\begin{align}
\Delta\ln T_{\rm nuc}
&\sim\left|\frac{d\ln J}{d\ln T}\right|^{-1}\notag\\
&\simeq6.35\times10^{-3}
\left(\frac{\mu}{10}\right)^{-3}
\left(\frac{s}{4}\right)^3
\left(\frac{q}{40}\right)^{-1}\notag\\
&\quad\times\left[\frac{|1-3s/(2q)|}{0.85}\right]^{-1}.
\label{eq:app_rise_width}
\end{align}
This estimate applies on the rising side of the burst, where the
slope varies slowly over the interval, rather than at the maximum
where the first derivative vanishes.

Since $\tau_{\rm cool}=|d\ln T/dt|^{-1}$, the associated interval is
\begin{align}
\Delta t_{\rm coll}
\simeq0.635\,{\rm day}
\left(\frac{\tau_{\rm cool}}{100\,{\rm day}}\right)
\left(\frac{\Delta\ln T_{\rm nuc}}{6.35\times10^{-3}}\right).
\end{align}
Under the short nucleation-and-depletion window approximation above,
\begin{align}
&f_{\rm grow}
\simeq6.35\times10^{-3}
\left(\frac{\mu}{10}\right)^{-3}
\left(\frac{s}{4}\right)^3
\left(\frac{q}{40}\right)^{-1}
\times\left[\frac{|1-3s/(2q)|}{0.85}\right]^{-1}.
\end{align}
This is an order-of-magnitude estimate of the effective interval,
not an identification of the microscopic growth and nucleation timescales.

This estimate can be compared with the normalization of the empirical
grain-size relation in the main text,
\begin{align}
r_{\rm gr}
&\simeq0.68\,{\rm \mu m}
\left(\frac{X'_{\rm C}\rho_{2000}\tau_{\rm cool,2000}}
{10^{-11}\,{\rm g\,cm^{-3}\,day}}\right).
\end{align}
On the other hand, using the full cooling time as the collision-active
duration at $T=2000\,{\rm K}$ gives
\begin{align}
r_{\rm gr}-a_{1{\rm C}}
&\simeq1.77\times10^2\,{\rm \mu m}\,\alpha_{\rm C}
\left(\frac{a_{1{\rm C}}}{1.281\times10^{-8}\,{\rm cm}}\right)^3\notag\\
&\quad\times\left(\frac{X'_{\rm C}\rho_{2000}\tau_{\rm cool,2000}}
{10^{-11}\,{\rm g\,cm^{-3}\,day}}\right).
\label{eq:app_full_growth}
\end{align}
For grains much larger than the monomer, the ratio of these normalizations gives
\begin{align}
f_{\rm grow,eff}
&\simeq3.83\times10^{-3}\,\alpha_{\rm C}^{-1}
\left(\frac{a_{1{\rm C}}}{1.281\times10^{-8}\,{\rm cm}}\right)^{-3}.
\label{eq:app_fgrow_scaled}
\end{align}
This value is comparable to the nucleation-window estimate.
The smaller normalization of the numerical grain-size relation can
therefore be interpreted in terms of a short effective collision
interval around nucleation and depletion, rather than the full cooling time.

\bibliographystyle{apj} 
\bibliography{sn_dust}

\end{document}